\documentclass[useAMS,usenatbib]{mn2e}

\usepackage{amsmath, amssymb}
\usepackage{epsfig}         
\usepackage{graphicx, float}
\usepackage{multirow}

\usepackage{natbib}
\usepackage{aas_macros}

\usepackage{url}
\usepackage[colorlinks=true, pdfstartview=FitV, linkcolor=blue, citecolor=blue, urlcolor=blue]{hyperref}

\def\Mpch{~h^{-1} {\rm Mpc}}
\def\MpchVolume{~(h^{-1} {\rm Mpc})^3}
\def\kpch{~h^{-1} {\rm kpc}}
\def\Msolar{~h^{-1} \rm{M}_{\odot}}

\newcommand{\spider}{NEXUS~}
\newcommand{\Spider}{NEXUS+~}
\newcommand{\spiderDen}{NEXUS\_den~}
\newcommand{\spiderTidal}{NEXUS\_tidal~}
\newcommand{\spiderDenlog}{NEXUS\_denlog~}
\newcommand{\spiderVeldiv}{NEXUS\_veldiv~}
\newcommand{\spiderVelshear}{NEXUS\_velshear~}
\newcommand{\logFilter}{Log-Gaussian~}

\newcommand{\SpideR}{NEXUS+}
\newcommand{\spiderDeN}{NEXUS\_den}
\newcommand{\spiderTidaL}{NEXUS\_tidal}
\newcommand{\spiderDenloG}{NEXUS\_denlog}
\newcommand{\spiderVeldiV}{NEXUS\_veldiv}
\newcommand{\spiderVelsheaR}{NEXUS\_velshear}

\newcommand{\Vector}[1]{\mathbf{#1}}

\newcommand {\lcdm}{$\Lambda$CDM~}

\newcommand{\eq}[1]{Eq. \eqref{#1}}
\newcommand{\eqs}[2]{Eqs. \eqref{#1}-\eqref{#2}}
\newcommand{\refsec}[1]{section \ref{#1}}
\newcommand{\reftab}[1]{Table \ref{#1}}
\newcommand{\reffig}[1]{Figure \ref{#1}}

\newcommand{\figDir}{small_figures_pdf/}

\newcommand{\MC}[1]{#1}

\newcommand{\hahn}{HPCD~}

\title[NEXUS: Tracing the Cosmic Web Connection]
{NEXUS: Tracing the Cosmic Web Connection}

\author[Marius Cautun, Rien van de Weygaert, Bernard J. T. Jones ]
       {Marius Cautun$^{1}$\thanks{E-mail : cautun@astro.rug.nl},
        Rien van de Weygaert$^{1}$ and
        Bernard J. T. Jones$^{1}$\\
$^1$  Kapteyn Astronomical Institute, University of Groningen, P.O. Box 800, 9747 AV Groningen, The Netherlands \\
}

\begin{document}


\maketitle

\begin{abstract}
We introduce the NEXUS algorithm for the identification of Cosmic Web environments: clusters, filaments, walls and voids. This is a multiscale and automatic morphological analysis tool that identifies all the cosmic structures in a scale free way, without preference for a certain size or shape. We \MC{develop} the NEXUS method to incorporate the density, tidal field, velocity divergence and velocity shear as tracers of the Cosmic Web. We also present the NEXUS+ procedure which, taking advantage of a novel filtering of the density in logarithmic space, is very successful at identifying the filament and wall environments in a robust and natural way.

To asses the algorithms we apply them to an N-body simulation. We find that all methods correctly identify the most prominent filaments and walls, while there are differences in the detection of the more tenuous structures. In general, the structures traced by the density and tidal fields are clumpier and more rugged than those present in the velocity divergence and velocity shear fields. We find that the NEXUS+ method captures much better the filamentary and wall networks and is successful in detecting even the fainter structures. We also confirm the efficiency of our methods by examining the dark matter particle and halo distributions. 
\end{abstract}


\section{Introduction}
\label{sec:introduction}
Early attempts to map the large scale distribution of galaxies in the universe \citep{Gregory78,Geller89,Lapparent86,Shectman96} revealed that galaxies are far from being evenly distributed across the nearby Universe. On the contrary, the mass distribution delineated by galaxies seems to form an intricate network of compact and dense associations interconnected by tenuous ``bridges" or ``filaments" surrounded by surprisingly vast empty regions \citep{Kirshner81}. Preliminary studies suggested that the universe on large scales could be described as a cellular system \citep{Joeveer78} or a \textit{Cosmic Web} \citep{Bondweb96}. This has been confirmed in recent times by large galaxy surveys such as the 2dFGRS \citep{Colless03}, the Sloan Digital Sky Survey \citep[e.g.][]{Tegmark04} and the 2MASS redshift survey \citep{Huchra05}. 

The Cosmic Web can be seen as the most prominent manifestation of the anisotropic nature of gravitational collapse, the motor behind the formation of structure in the cosmos \citep{Peebles80}. N-body computer simulations have illustrated how a primordial field of tiny Gaussian density perturbations transforms into a pronounced and intricate filigree of filamentary features, dented by dense compact clumps at the nodes of the network \citep[][]{Jenkins98,Colberg05,Springel05b,Dolag06}. The description of the Megaparsec matter distribution as an interconnected network or a cosmic web is not a coincidence. Even early computer simulations indicated the close connection between each morphological component, namely that clusters sit at the intersection of filaments and filaments are formed at the intersection of walls \citep{Doroshkevich80,Melott83,Pauls95,Shapiro83,Sathyaprakash96}.

One of the main reasons for our interest in outlining the Cosmic Web concerns the question whether and how far the weblike environment influences the properties and evolution of galaxies. Recent $N$-body simulations have found that the filamentary or sheetlike nature of the environment has a distinct influence on the shape and spin orientation of dark matter haloes \citep{Aragon07a,Hahn2007a,Hahn2007b,Paz08,Hahn09,HahnPhd09,Zhang09}. Other recent works \citep[][]{Jones10,2012arXiv1207.0068T} have shown that indeed there is an alignment, even though weak, of galaxies and the filaments they lie within. In this paper we propose new robust and flexible methods that allow a better identification of the Cosmic Web environments and hence help us to better understand how environments influence the formation and evolution of dark matter haloes and galaxies.

\subsection{Cosmic Web Detection}
Identifying the components of the Cosmic Web is a major challenge due to the overwhelming complexity of the individual structures as well as their connectivity, the lack of structural symmetries, its intrinsic multiscale nature and the wide range of densities found in the cosmic matter distribution. Over the years, a variety of heuristic measures were forwarded to analyse specific aspects of the spatial patterns in the large scale Universe, but only recently these have lead to a more solid and well-defined machinery for identifying the Cosmic Web. Nearly without exception, these methods borrow extensively from other branches of science such as image processing, mathematical morphology, computational geometry and medical imaging.

The connectedness of elongated supercluster structures in the cosmic matter distribution was first probed by means 
of percolation analysis, introduced and emphasized by Zel'dovich and coworkers \citep{Zeldovich82,Shandzeld89,Shandarin04,Shandarin09b}, while a related graph-theoretical construct, the minimum spanning tree of the galaxy distribution, was extensively analysed by Bhavsar and collaborators \citep{Barrow85,Graham95,Colberg07} in an attempt to develop an objective measure of filamentarity. Both \citet{Colberg05} and by \citet{Pimbblet05} set out to identify filaments and their adjoining clusters, using quite different techniques. 

More general filament finders have been put forward by a number of authors. Following specific physical criteria, 
\cite{Gonzalez09} recently forwarded an interesting and promising combination of a tessellation-based density estimator 
and a dynamical binding energy criterion. A thorough mathematical nonparametric formalism involving the medial axis of 
a point cloud, as yet for 2-D point distributions, was proposed by \cite{Genovese10}. It is based on a geometric representation of filaments as the medial axis of the data distribution. Also solidly rooted within a geometric 
and mathematical context is the more generic geometric inference formalism developed by \cite{Chazal09}. It allows the 
recovery of geometric and topological features of the supposedly underlying density field from a sampled point cloud 
on the basis of distance functions. \citet{Stoica05,Stoica07,Stoica10} use a generalization of the classical Candy model to locate and catalogue filaments in galaxy surveys. This approach has the advantage that it works directly with the original point process and does not require the creation of a continuous density field. However, computationally it is very demanding. 

The more recent formalisms that are intent on characterizing the full range of weblike formalisms usually exploit the 
morphological information in the gradient and Hessian of the density field or potential field, i.e. the tidal field \citep[see e.g.][]{Aragon07a,Aragon07b,Hahn2007a,Hahn2007b,Sousbie08a,Forero-Romero09,Bond09,Bond10}. Morse theory \citep[see][]{Colombi2000} forms the basis of the {\it skeleton analysis} by \cite{Novikov06} (2-D) and \cite{Sousbie08a} (3-D). It identifies morphological features with the maxima and saddle points in the density field and results in an elegant and mathematically rigorous tool for filament identification. However, it is computationally intensive, focuses mostly on filaments and is strongly dependent on the smoothing scale of the density field. A more elaborate classification scheme on the basis of the manifolds in the tidal field -- involving all morphological features in the cosmic matter distribution -- has been forwarded by \cite{Hahn2007a} \citep[also see][]{Hahn2007b,Forero-Romero09}.

Instead of using the tidal field configuration, one may also try to link directly to the morphology of the density field itself. Though this allows a more detailed view of the intricacies of the multiscale matter distribution, it is usually more sensitive to noise and less directly coupled to the underlying dynamics of structure formation than the tidal field morphology. A single scale dissection of the density field into its various morphological components has been done by \cite{Bond09}, and applied to N-body simulations and galaxy redshift samples \citep[also see][]{Bond10,Choi10}. A more elaborate formalism is the Multiscale Morphology Filter (MMF), introduced by \cite{Aragon07b}. It looks at structure from a scale-space point of view, treating the spatial structure in $D$ dimensions in the context of an explicit $D+1$ dimensional space \citep{Florack92,Lindeberg98}. The $D+1$ dimensional scale space consists of the $D$-dimensional spatial structure at a range of spatial resolution scales. The MMF formalism subsequently assigns a local morphology based on an evaluation of the multiscale second order variations in the local density field. Instead of restricting the analysis to one particular scale, by evaluating the density field Hessian over a range of spatial scales and determining at which scales and locations the various morphological signatures are most prominent, the MMF explicitly addresses the multiscale nature of the cosmic structures. A somewhat similar multiscale approach was followed by the Metric Space Technique described by \cite{Wu09}, who applied it to a morphological analysis of the DR5 of the SDSS. 

A more recent development is that of the Spineweb procedure \citep{Aragon10}, which traces the various features of the cosmic web on pure topological grounds by invoking the {\it Watershed Transform} (WT). \MC{The WT was introduced by \cite{Platen07} as the basis of the Watershed Void Finder technique \citep{Platen07} which identifies cosmic voids with the watershed basins.} The Spineweb procedure elaborates on this, by identifying the central axis of filaments and the inner plane of walls with the boundaries between the watershed segments of the density field. While the basic Spineweb procedure involved one scale, the full procedure allows a multiscale topological characterization of the Cosmic Web \cite{Aragon10b}. However, to do so it must invoke some implicit assumptions on the connectivity of the various topological features.

\subsection{Intention and Outline}
The goal of this paper is to present two new algorithms (\spider and \SpideR) for the detection of Cosmic Web environments and to assess their effectiveness. Elaborating on the multiscale scale-space context of the rudimentary density field Multiscale Morphology Filter (MMF, \cite{Aragon07}), the \spider and \Spider formalism represents a complete and versatile instrument for the structural and physical study of the Cosmic Web. Simultaneously taking into account the multiscale nature of the cosmic mass distribution, \spider and \Spider explicitly operate \MC{on a diversity of physical fields that are relevant to the formation and evolution of the Cosmic Web.}
The extension beyond the density field, towards the use of information contained in the tidal field, velocity divergence and velocity shear to trace the large scale structure, is a key aspect of \spider and \SpideR. The new formalism allows us to compare the environments traced by both the positional and velocity part of the phase space. We focus most of the analysis on the detection of filaments and walls, since these are the most challenging environments to identify. We find that our new methods are very efficient at tracing the Cosmic Web, resulting in very high quality filaments and walls. 

This study is organized as follows. In sections~\ref{sec:spider} and \ref{subsec:Spider_description} we describe the \spider and \Spider methods which we use for the identification of the Cosmic Web, including a comparison of the two algorithms on a toy model. This is followed in \refsec{sec:spider_extension} with an extension of the tools to use a multitude of cosmological fields (density and tidal field versus the velocity divergence and velocity shear) as tracers of the cosmic environments. This way we take full advantage of the full 6-D information contained in phase space.

The second part of the paper is focused on assessing how these methods cope with the complex and hierarchical structures present in the universe. To do so we use the DTFE density and velocity divergence from N-body simulations as inputs to our algorithms -- see \refsec{sec:nbody}. Section~\ref{sec:environments} presents the cluster, filament and wall environments identified in the simulation and compares the results of the different methods. To asses the quality of the detections we look at the dark matter particle and halo distributions in each environment and also study the effects of a multiscale versus single scale approach in tracing the Cosmic Web. Finally, we summarize our findings in \refsec{sec:conclusions}.

\section{\spider: general formalism for multiscale morphological analysis}
\label{sec:spider}
The \spider algorithm is a Scale-Space method for morphologically segmenting the Cosmic Web into its three distinct features: clusters, filaments and walls. The environment identification is performed in a scale and user independent way to account for the multiscale nature of the large scale structure, which is the result of the hierarchical evolution of the cosmic mass distribution. The method is derived from the field of medical imaging \citep{DBLP:conf/miccai/FrangiNVV98,Sato983dmulti-scale,li:2040} where it is used to identify nodules, vessels and walls in two- and three-dimensional images. An earlier and simpler version of the method was introduced in cosmology under the name Multiscale Morphology Filter (MMF) in \citet{Aragon07b}. The MMF involved a rudimentary treatment of scale-space analysis and restricted itself to the use of the density field as tracer of the Cosmic web environment.

In general, the scale-space formalism can be applied to any input field to detect point-, line- and sheetlike structures in the field values, and thus lends itself to applications involving a range of quantities dynamically relevant for the formation and evolution of the Cosmic Web. Following this observation, we have embedded the scale-space formalism in the physical framework of Cosmic Web formation. 

For this purpose we have defined two classes of the algorithm, the \spider and the \Spider formalism. The main difference between \spider and \Spider concerns the filter used for constructing the representation of the field at different resolutions in scale-space. The \spider technique uses a Gaussian filter for smoothing while \Spider uses a \logFilter filter (more on that in \refsec{subsec:Spider_description}). We will demonstrate in this study that they yield a substantially more realistic and robust representation of filaments, walls and their mutual connectivity, over the range of scales covered by the scale-space representation.

\subsection{\spider: general algorithm description}
\label{subsec:spider_description}
The \spider algorithm detects the point-, line- and sheet-like structures\footnote{The algorithm given here applies to detecting the point-, line- and sheet-like structures corresponding to maxima in the field values. If we are interested in the same structures but for the minima of the field values, than we need to apply the same algorithm to $-f$. For example the Cosmic Web environments correspond to maxima in density but to minima in velocity divergence.} for a generic input field $f$. For large scale structure, these features correspond to clusters, filaments and walls. To keep the notations clear, we limit our discussion to the Cosmic Web environments, but there is no loss of generality. The \spider algorithm consists of the following six steps:
\begin{list}{}{\leftmargin=3em \rightmargin=0cm \labelwidth=3em}
    \item[(I)] Applying a Gaussian filter of width $R_n$ to the input field.
    \item[(II)] Computing the Hessian matrix eigenvalues for the filtered field.
    \item[(III)] Assigning to each point a cluster, filament and wall signature using the Hessian eigenvalues .
    \item[(IV)] Repeating steps \textit{(I)} to \textit{(III)} over a range of smoothing scales $(R_0,R_1,..,R_N)$, to 
construct the scale-space representation of the field.
    \item[(V)] Combining the results of all scales to obtain a scale independent cluster, filament and wall signature.
    \item[(VI)] Using physical criteria to determine the detection threshold corresponding to valid environments.
\end{list}
In the following we elaborate on each step of the algorithm and give the details necessary for the implementation of the method.
\begin{figure}
    \centering
    \includegraphics[width=0.9\linewidth,angle=0.0]{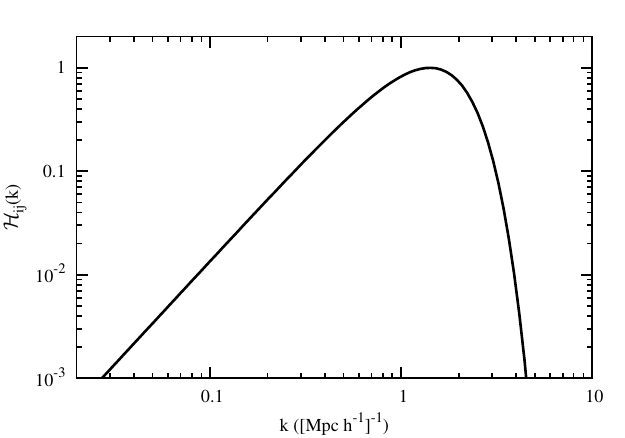}
    \caption{The amplitude of the kernel function $\mathcal{H}_{ij}$ for a smoothing radius of $1\Mpch$. The values were normalized such that $\mathcal{H}_{ij}$ has a maximum value of 1. Note the logarithmic axes.}
    \label{fig:hessian_kernel}
\end{figure}

\subsubsection{Step I: Applying Gaussian smoothing}
A Gaussian filter of width $R_n$ is applied to the input field $f(\Vector{x})$. This gives rise to a smoothed field $f_{R_n}(\Vector{x})$ given by:
\begin{equation}
    f_{R_n}(\Vector{x}) = \int \frac{d^3k}{(2\pi)^3} e^{-k^2R_n^2/2} \hat{f}(\Vector{k})  e^{i\Vector{k}\cdot\Vector{x}} ,
    \label{eq:filtered_field}
\end{equation}
where $\hat{f}(\Vector{k})$ is the Fourier transform of the input field $f(\Vector{x})$.

\subsubsection{Step II: Computing Hessian eigenvalues}
The Hessian of the filtered field is computed as:
\begin{equation}
    H_{ij,R_n}(\Vector{x}) = R_n^2 \; \frac{\partial^2f_{R_n}(\Vector{x})}{\partial x_i\partial x_j},
    \label{eq:hessian_general}
\end{equation}
where $ H_{ij,R_n}$ represents the $i,j$ entry of the $H_{R_n}$ Hessian matrix. The $R_n^2$ term is a renormalization factor that has to do with the multiscale nature of the \spider algorithm. It makes sure that the same weight is assigned when comparing the Hessian value at different scales. Using \eq{eq:filtered_field}, the Fourier transform of the Hessian reads:
\begin{equation}
    \hat{H}_{ij,R_n}(\Vector{k}) = \mathcal{H}_{ij,R_n}(\Vector{k}) \; \hat{f}(\Vector{k})
    \label{eq:hessian_fourier}
\end{equation}
with $\mathcal{H}$ the Hessian kernel function given by:
\begin{equation}
    \mathcal{H}_{ij,R_n}(\Vector{k}) = -k_ik_j R_n^2 e^{-k^2R_n^2/2}.
    \label{eq:hessian_kernel}
\end{equation}
The kernel function characterizes which Fourier components of the input field give contributions to the Hessian matrix. The dependence of the Hessian kernel on $k$ is shown in \reffig{fig:hessian_kernel}. At a given smoothing scale $R_n$, the $\mathcal{H}$ kernel has a peak at $k_\rmn{peak}=\sqrt{2}/{R_n}$, with a sharp drop-off for higher $k$ and a linear fall for smaller $k$. Therefore, for a given scale $R_n$, only the Fourier components of $f$ around the peak $k_\rmn{peak}$ will give an important contribution to the Hessian matrix.

The \spider formalism depends only on the eigenvalues of the Hessian matrix, eigenvalues given by:
\begin{equation}
    \det( H_{R_n}(\Vector{x})-\lambda_{a,R_n}(\Vector{x}) ) = 0, \;\rmn{with}\; \lambda_1 \le \lambda_2 \le \lambda_3. 
    \label{eq:hessian_eigenvalues}
\end{equation}

\subsubsection{Step III: Computing environment signature}
The eigenvalues of the Hessian matrix can be used to assign a cluster, filament and wall characteristic to every point $\Vector{x}$ using the expected behaviour given in \reftab{tab:eigenvalues}. This is the environment signature and is denoted with $\mathcal{S}(\Vector{x})$.
\begin{table}
    \caption{Hessian eigenvalue relationships for the different environments of the Cosmic Web. The second column gives the qualitative relationships between the eigenvalues (conditions that are implemented analytically in \eq{eq:shape_strength} ) while the third column gives strict eigenvalues constraints implemented in \eq{eq:response}.}
    \label{tab:eigenvalues}
    \begin{tabular}{lll}
        \hline
        Structure & Soft constraints & Strict constraints \\
        \hline
        cluster & $|\lambda_1|\simeq|\lambda_2|\simeq|\lambda_3|$ & $\lambda_1<0$; $\lambda_2<0$; $\lambda_3<0$ \\
        filament & $|\lambda_1|\simeq|\lambda_2|\gg|\lambda_3|$ & $\lambda_1<0$; $\lambda_2<0$ \\
        wall & $|\lambda_1|\gg|\lambda_2|$; $|\lambda_1|\gg|\lambda_3|$ & $\lambda_1<0$ \\
        \hline
    \end{tabular}
\end{table}
The first step in computing the signature is to define the shape strength $\mathcal{I}$. This gives a quantitative description of the approximate relations given in the middle column of \reftab{tab:eigenvalues}. The shape strength is defined as:
\begin{equation}
    \mathcal{I} = 
    \begin{cases}
        \left| \frac{\lambda_{3}}{\lambda_{1}} \right| & \rmn{cluster}  \vspace{.1cm} \\
        \left| \frac{\lambda_{2}}{\lambda_{1}} \right| \Theta\left( 1-\left| \frac{\lambda_{3}}{\lambda_{1}} \right|\right) & \rmn{filament}  \vspace{.1cm} \\
        \Theta \left( 1-\left| \frac{\lambda_{2}}{\lambda_{1}} \right|\right) \Theta\left( 1-\left| \frac{\lambda_{3}}{\lambda_{1}} \right|\right) & \rmn{wall}
    \end{cases}
    \label{eq:shape_strength}
\end{equation}
where we use the notation $\Theta(x)=x\theta(x)$ for clarity, with $\theta(x)$ the step function ($\theta(x)=1$ if $x\geq0$, $0$ otherwise). The strength $\mathcal{I}$ is large when the eigenvalues at $\Vector{x}$ correspond to a prominent structure and small otherwise. The cluster/filament/wall signature is defined as:
\begin{equation}
    \mathcal{S} = \mathcal{I} \times
    \begin{cases}
        |\lambda_{3}| \; \theta(-\lambda_{1}) \theta(-\lambda_{2}) \theta(-\lambda_{3}) & \rmn{cluster} \\
        |\lambda_{2}| \; \theta(-\lambda_{1}) \theta(-\lambda_{2}) & \rmn{filament} \\
        |\lambda_{1}| \; \theta(-\lambda_{1}) & \rmn{wall,}
    \end{cases}
    \label{eq:response}
\end{equation}
where the $\theta(-\lambda_a)$ factors (with $a=1,2,3$) incorporate the right most column of \reftab{tab:eigenvalues}. The $|\lambda_{a}|$ term gives the intensity of the morphological feature and can be used to discriminate between real signals (large $|\lambda_{a}|$) and noise (small $|\lambda_{a}|$).

\subsubsection{Step IV: Computing the environmental signature over a range of smoothing scales}
The previous three steps are repeated over a range of smoothing scales $(R_0,R_1,..,R_N)$. The hierarchy of smoothing scales is taken as $R_n=(\sqrt{2})^nR_0$ with $R_0$ the smallest scale at which one expects to find structures \citep{Sato983dmulti-scale}. Taking an even smaller step between any two successive smoothing scales makes only minor differences. In practice we choose $R_0$ equal to the grid spacing of the input field. We found that for the detection \MC{of the most prominent features of the} Cosmic Web it is sufficient to consider smoothing scales in the range $0.5\Mpch$ to $4\Mpch$. \MC{In this respect it is good to note that outstanding features of the Cosmic Web are visible within a particular range of scales centred around the transition scale between linear and non-linear structures.}

The result of this step is a signature function for each scale $\mathcal{S}_{R_n}(\Vector{x})$ which characterizes the environmental response of point $\Vector{x}$ at the $R_n$ smoothing scale.

\subsubsection{Step V: Scale-space stacking}
The signature of the given set of scales is combined to obtain the overall signature. This is a scale independent map characterizing the degree to which the point $\Vector{x}$ is part of a cluster, filament or wall. A structure of a given size will give the largest signature for a smoothing scale of the same size. Therefore, the overall signature at a point is the maximum signature over all the scales:
\begin{equation}
    \mathcal{S}(\Vector{x}) = \max_{\rmn{levels\;}n} \mathcal{S}_{R_n}(\Vector{x}).
    \label{eq:total_response}
\end{equation}

\subsubsection{Step VI: Computing the detection threshold}
\begin{figure}
    \centering
    \includegraphics[width=.97\linewidth]{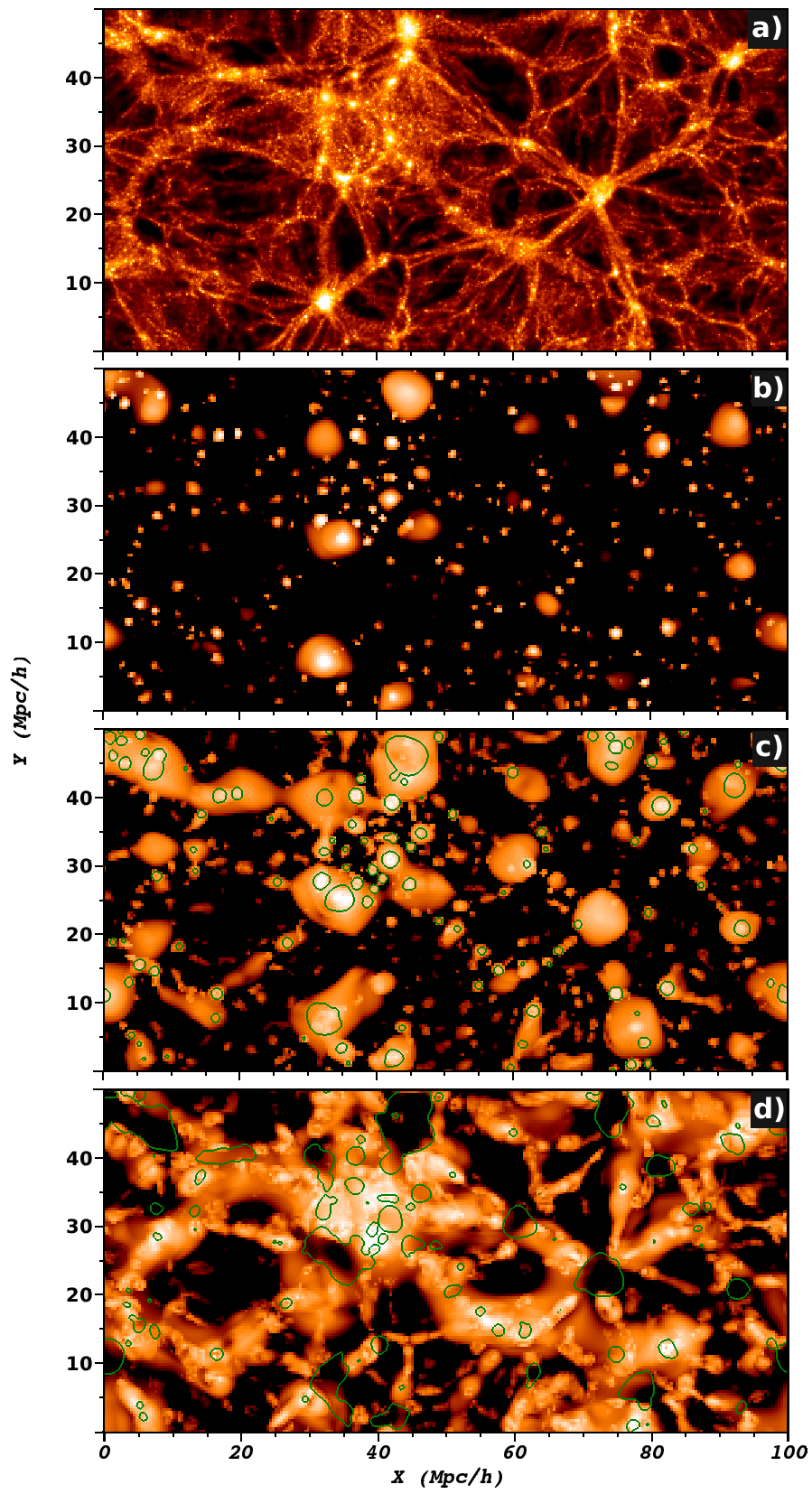}
    \caption{A $1\Mpch$ slice through: a) density field, b) node signature, c) filament signature and d) wall signature obtained from an N-body simulation. The white, orange and black show the high, medium and low values of density and environment signature. The green line contours show the regions with high node signature (panel c) and with high filament signature (panel d).}
    \label{fig:signature}
\end{figure}
The signature has a wide range of values, with the large one corresponding to strong structures and the small ones coming from noise and null detections. This can be appreciated in \reffig{fig:signature}, which shows the cluster, filament and wall signature. Therefore, the last step in the algorithm involves the use of physical criteria to find the threshold signature that discriminates between valid and invalid detections. Signature values larger than the threshold correspond to real structures while the rest are spurious detections. The threshold signature for clusters is found by requiring that the identified objects are virialized, whereas for filaments and walls the threshold is given by the dependence of the filament/wall mass with environmental signature.

\begin{figure}
    \centering
    $\begin{array}{c}
    \includegraphics[width=.65\linewidth,angle=-90.0]{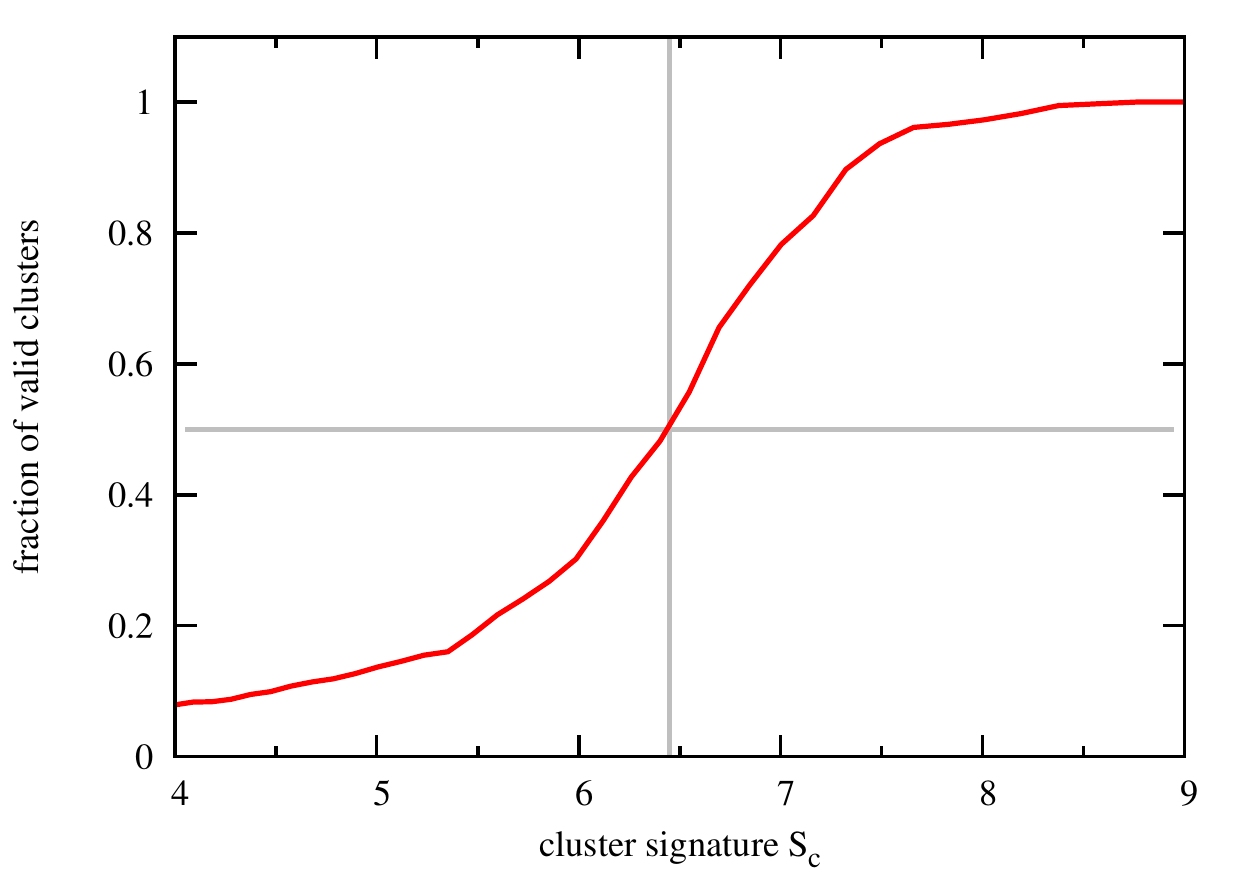} \\
    \includegraphics[width=.65\linewidth,angle=-90.0]{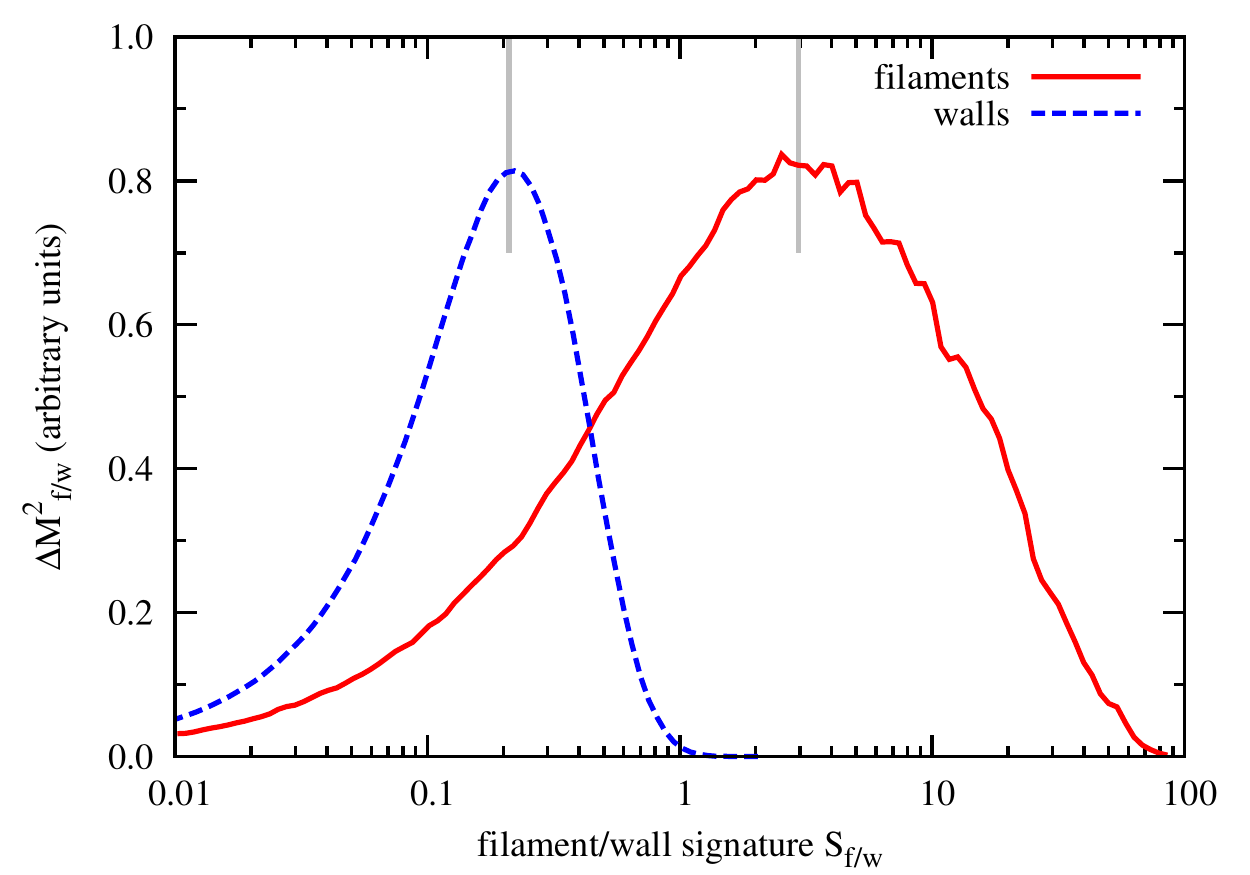}
    \end{array}$
    \caption{\textit{Upper panel}: The dependence of the fraction of clusters with density larger than the virial density versus the cluster signature $\mathcal S_c$. The intersection of the gray lines shows the cluster detection threshold. \textit{Lower panel}: Determination of the detection threshold for filaments and walls. The peak of $\Delta M^2$ (shown by the gray vertical lines) corresponds to the signature threshold for filament and wall identification (see text for details).}
    \label{fig:optimalThreshold}
\end{figure}
The procedure to determine the signature threshold for cluster detection is illustrated in the upper panel of \reffig{fig:optimalThreshold}. Clusters are the largest and most recently formed fully virialized objects \citep{2005RvMP...77..207V}. We use this definition to determine the signature threshold for cluster identification. We test for virialization by requiring that the average density of the cluster is larger than $\Delta=370$, which is the value given by the spherical collapse model at $z=0$ \citep{1972ApJ...176....1G}. From \reffig{fig:optimalThreshold} it can be seen that the fraction of objects with an average density larger than the virialization threshold changes very fast from 0 to 1 as we increase the cluster signature $\mathcal{S}_c$. We then take the signature threshold as the value where half of the objects have a density larger than $\Delta$.

The filament and wall identification is performed by limiting our detections to only the most prominent filamentary and wall regions. We find that the same method can be successfully used for the recognition of both filaments and walls. Let us denote with $M_f(\mathcal{S}_f)$ the mass in filaments with a signature value larger or equal to $\mathcal S_f$. As $\mathcal S_f$ decreases, more and more regions are included and hence $M_f(\mathcal{S}_f)$ increases. Most of the change in this function is restricted to a small range in $S_f$ values and it gives a natural way of discriminating between real and spurious detections. In Appendix \ref{sec:optimal_detection_threshold} we show that the mass change with signature:
\begin{equation}
    \Delta M_f^2 = \left| \frac{dM_f^2}{d\log\mathcal{S}_f} \right|
    \label{eq:delta_M2_filament}
\end{equation}
gives a natural and robust method of defining the most prominent filamentary components of the Cosmic Web. Similarly, in the case of walls, we can define $\Delta M_w^2$ using the above equation with $M_f$ and $S_f$ replaced by their corresponding quantities for wall environments, $M_w$ and $S_w$. The quantity $M_w(\mathcal{S}_w)$ is the mass in regions that have a wall signature value larger or equal to $\mathcal S_w$. 
The $\Delta M^2$ dependence for both filaments and walls is shown in \reffig{fig:optimalThreshold}. We use the pronounced $\Delta M^2$ peak to delineate the valid environments, which are the points with signatures larger than the position of the $\Delta M^2$ peak. All other points with smaller signature are considered null detections. This threshold method reproduces very well the filamentary and wall network visible in both the cosmic density and velocity divergence fields.

The algorithm performs the environment detection by applying the above steps first to clusters, then to filaments and finally to walls. This sequence (first clusters, then filaments and finally walls) has to be followed due to presence of anisotropic clusters and filaments that give mixed environmental signatures. This can be appreciated from panel c of \reffig{fig:signature} where on top of the filament signature we show the contours corresponding to large cluster signature. We see that there are several regions that have both a large cluster and filamentary characteristic. This is due to non-spherical clusters that have a large filamentary signature. Similar anisotropic cluster/filaments may give a strong wall signature (see panel d of \reffig{fig:signature}). To overcome this cross-contamination, a point is part of a filament only if it was not previously identified as in a cluster. Similarly a point is in a wall if it was not previously identified as part of a cluster or filament. This procedure makes sure that each point is assigned a single classification: cluster, filament, wall or field (everything else that is not a cluster, filament or wall).

\section{\Spider: logarithmic formalism for multiscale morphological analysis}
\label{subsec:Spider_description}

\begin{figure}
    \centering
    \includegraphics[width=\linewidth]{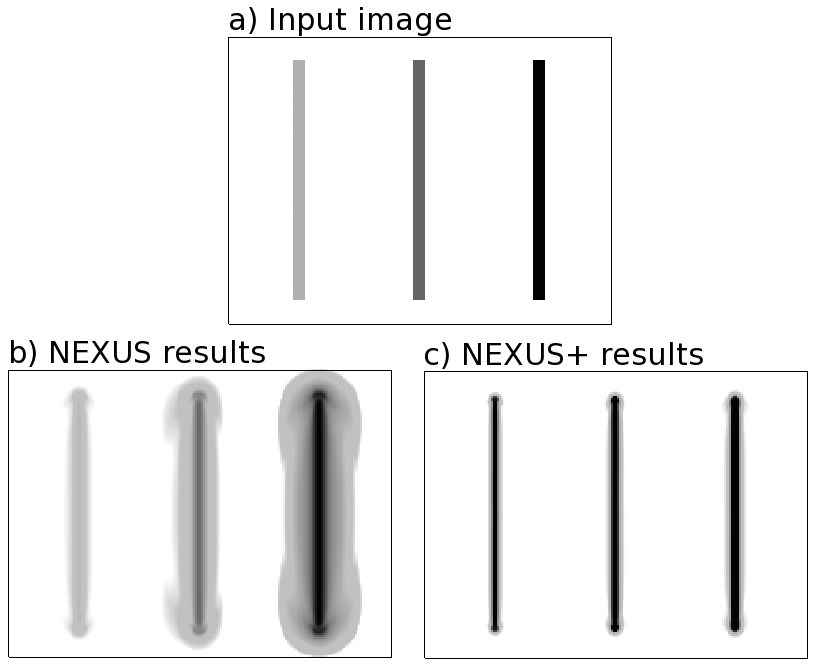}
    \caption{The filaments detected using the \spider and \Spider methods when applied to a test image. The input test image (upper panel) contains three filaments with the same width but of different intensities: 1, 10 and 100 (from left to right). Frame b) shows the \spider filament signature with a threshold low enough such that also the weakest filament (left most one) is visible. Panel c) depicts the same as frame b) but for the \Spider method.}
    \label{fig:toy_model}
\end{figure}

The \spider algorithm is very efficient in detecting the environments in a field $f$ where all the structures correspond to the same order of magnitude values of $f$. However, the method faces some challenges when the structures in $f$ are present over orders of magnitude in field values. To better understand this, we present a test example in \reffig{fig:toy_model}. It shows three filaments characterized by different intensities: 1, 10 and 100 (from left to right). For the \spider method to identify all the three filaments, the threshold needs to be so low that the stronger filaments are detected as extending much beyond their input data boundaries. The higher intensity peaks give a significant signal even at large distances, due to the combination of the Gaussian filter not dropping off fast enough and the high field value of the peak.

One way to remedy this problem is to replace the Gaussian filter with a new smoothing method that takes into account the large range in values. For that we introduce the \logFilter filter, which is a Gaussian filter in logarithm space. By replacing the Gaussian filter in \spider with the \logFilter filter we obtain the \Spider formalism. The results of the new method are presented in frame c of \reffig{fig:toy_model}. It clearly shows that the new filtering procedure recovers much better the three filaments.

\subsection{\Spider: general algorithm description}
The main difference between \spider and \Spider is the use of the \logFilter filter instead of the Gaussian one. The steps of the \Spider algorithm are the same as the steps of \spider with the exception of steps \textit{(I)} and \textit{(II)}.

\subsubsection{Step I: Applying \logFilter smoothing}
A \logFilter filter of width $R_n$ is applied to the input field $f$. To this end, we introduce the field $g$, the logarithm of field $f$,
\begin{equation}
g = \log_{10} f\ ,
\end{equation}
\noindent and the field $g_{R_n}$, the smoothed logarithm at scale $R_n$, 
\begin{equation}
    g_{R_n}(\Vector{x}) = \int d^3y \; g(\Vector{y}) \; W_{G,R_n}(\Vector{x},\Vector{y}),
\end{equation}
with $W_{G,R_n}$ the Gaussian filter of width $R_n$.

Following the introduction of these quantities, the application of the \logFilter filter consists of three main steps, 
\begin{list}{}{\leftmargin=3em \rightmargin=0cm \labelwidth=2em}
    \item[1.] Computing the logarithm of the field $f$, $g=\log_{10}f$.
    \item[2.] Applying the Gaussian filter of width $R_n$ to $g$ to obtain the smoothed logarithm $g_{R_n}$. 
    \item[3.] Computing the smoothed field $f_{R_n}$ by taking the exponential of the smoothed logarithm $g_{R_n}$.
\end{list}

\noindent In practice, we perform the convolution of the field $g$ with the Gaussian filter in Fourier space, 
\begin{equation}
g_{R_n}(\Vector{x})=\int \frac{d^3k}{(2\pi)^3} e^{-k^2R_n^2/2} \hat{g}(\Vector{k})  e^{i\Vector{k}\cdot\Vector{x}},
\end{equation}
involving the simple multiplication of the Fourier field component $\hat{g}(\Vector{k})$ with 
the Gaussian exponential,
\begin{equation}
    \hat{g}_{R_n}(\Vector{k}) = e^{-k^2R_n^2/2} \; \hat{g}(\Vector{k}).
    \label{eq:Spider_fourier_multiplication}
\end{equation}

\noindent Subsequently, the resulting \Spider smoothed field $f_{R_n}$ is obtained by evaluating, 
\begin{equation}
    f_{R_n}(\Vector{x}) = C_{R_n} \; 10^{g_{R_n}}.
    \label{eq:Spider_smoothing}
\end{equation}
The variable $C_{R_n}$ is a multiplication constant that assures the mean of the input field is the same 
before and after filtering. 

\subsubsection{Step II: Computing Hessian eigenvalues}
The second step is the same as for the \spider algorithm, but since the smoothing filter is different some of the equations will also change. Now the Hessian of the smoothed field:
\begin{equation}
    H_{ij,R_n}(\Vector{x}) = R_n^2 \; \frac{\partial^2 f_{R_n}(\Vector{x})}{\partial x_i\partial x_j},
    \label{eq:Spider_hessian_general}
\end{equation}
can be written in Fourier space using:
\begin{equation}
    \hat{H}_{ij,R_n}(\Vector{k}) = -k_ik_j R_n^2 \hat{f}_{R_n}(\Vector{k}).
    \label{eq:Spider_hessian_fourier}
\end{equation}
Please note that now one cannot formulate $\hat{f}_{R_n}(\Vector{k})$ as a simple analytical expression as in the case of the \spider algorithm's \eq{eq:hessian_fourier}. For the \Spider algorithm one needs to perform steps \textit{(I)} and \textit{(II)} separately.

The rest of the steps are the same as \spider steps \textit{(III)} to \textit{(VI)} described in \refsec{subsec:spider_description}. It is important to note that because \Spider uses the logarithm of $f$ it can only be applied to input fields that have positive values at every point.

\subsection{\Spider on the density field}
\begin{figure}
    \centering
    \includegraphics[width=1\linewidth]{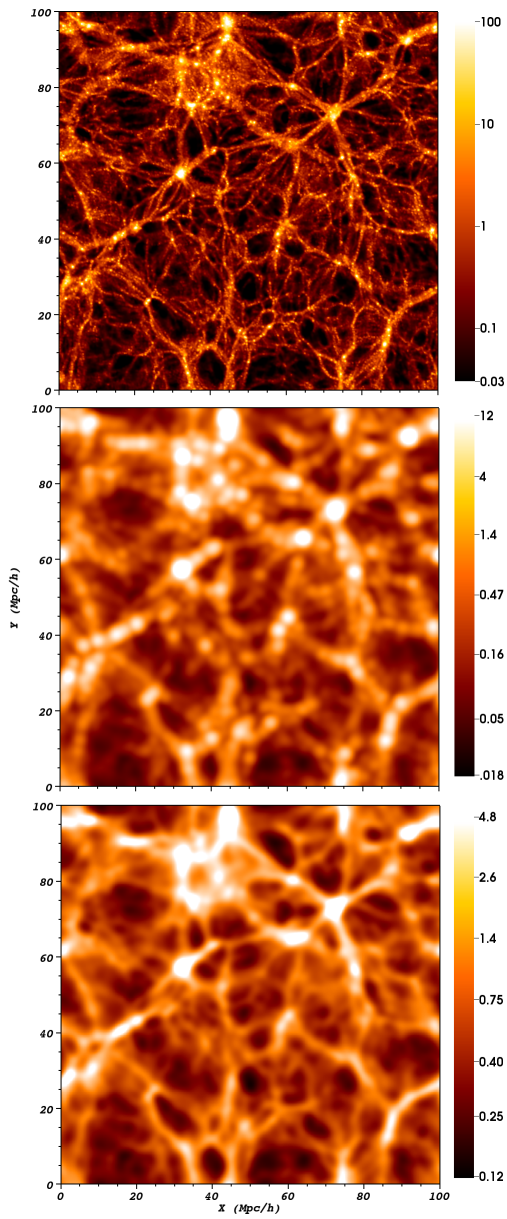}
    \caption{Comparing Gaussian and \logFilter density smoothing (see text for details). The upper panel shows the initial density field, while the central and lower panels give the Gaussian and respectively \logFilter smoothed density. Both cases were obtained using a $1\Mpch$ smoothing. The scale in the lower two panels was selected by fitting the density histogram with a log-normal distribution and plotting the values in the $peak-3\sigma$ to $peak+3\sigma$ range (with $peak$ and $\sigma$ the peak and standard deviation of the log-normal distribution).
    }
    \label{fig:logFilter_comparison}
\end{figure}
The major challenge of structure detection lies in the fact that the nonlinear density field, following its evolution, ranges over many orders of magnitude between the underdense and overdense regions. Structures are present over the whole range of values in density. To deal with this challenge we can use two approaches: either take the density logarithm (see \refsec{subsec:spider_den_log}) or use a different algorithm that takes into account the approximative log-normal shape of the density distribution. Here we take the former approach and apply the \Spider algorithm to the density field.

The strength of the \Spider algorithm can be easily appreciated if one compares the density field using Gaussian versus \logFilter smoothing. While there is a one to one mapping between density and density logarithm, this relation does not hold when one compares the smoothed density with the smoothed logarithm of the density. The results of the two methods are shown in \reffig{fig:logFilter_comparison}. We immediately observe that the Gaussian filtered density is dominated by several peaks with a typical spherical shape. On the other hand, the \logFilter results seem to trace much better the large scale structure. 

Most of the differences between the two results come from the very high density peaks. When applying a Gaussian smoothing, these higher density peaks gets smoothed up to large distances and dominate the signal coming from other less dense neighbouring regions. This leads to a loss of information about the large scale structure around these peaks. In the case of \logFilter smoothing, by taking the logarithm of the input field the contrast between these very high density peaks and their neighbourhoods is greatly reduced. Therefore the contribution of the peaks will not be dominant, even though the \logFilter filter has the same spatial extension as the Gaussian one.

\section{Tracer fields of the Cosmic Web: Extending the Nexus algorithm beyond density}
\label{sec:spider_extension}
There are various methods that attempt to identify the components of the Cosmic Web. These not only implement different detection techniques, but in many cases differ in the nature of the field used to trace the underlying cosmic structure. In other words, the variation in the results of different methods should not only be ascribed to the algorithms used, but also to the differences in the tracer fields. In this section we extend the \spider method to a multitude of Cosmic Web tracers: density, tidal field, velocity divergence and shear as well as to the density logarithm. By doing so we not only find the field with the best footprint of the Cosmic Web, but also gain better understanding of the evolution and structure of the Cosmic Web.

The most widely used tracers of the Cosmic Web is the density \citep{Aragon07b,2008ApJ...672L...1S,2011MNRAS.414..384S} and the tidal field \citep{Hahn2007a,Forero-Romero09}. On the other hand, the use of the density logarithm for the classification of the Cosmic Web is a novel method that we introduce here. Later on we will argue why this is a natural structure tracer field that one should consider. While there have been a fair share of methods using the positional information of the phase space, there are very few works that use the velocity field as Cosmic Web tracer. Most interesting is the work by \cite{2011JCAP...05..015S}, which emphasized the importance of the velocity field for understanding the emerging patterns in the matter distribution. Based on this, he used the variance of the velocity field as a measure of the local environment. In a sequence of studies that follows up on this idea \citep{Abel11,Shandarin12,Neyrinck12}, the full phase space structure of the mass distribution is used for an impressively accurate dynamical characterization of morphological structure of the Cosmic Web. Following this promising avenue, we are also working on relating the \spider and \Spider formalism to the structure found by these phase space based methods. 

\begin{table}
    \caption{The methods resulting from the extension of the \spider algorithm to several tracer fields of the Cosmic Web.}
    \label{tab:method_names}
    \begin{tabular}{ll}
        \hline
        Method name & Tracer field \\
        \hline
        \spiderDen     &  density field $\delta$ \\
        \spiderTidal   &  tidal field $ T $ \\
        \spiderDenlog  &  density logarithm field $\log_{10}(1+\delta)$ \\
        \spiderVeldiv  &  velocity divergence field $\theta$ \\
        \spiderVelshear & velocity shear field $\sigma$ \\
        \hline
    \end{tabular}
\end{table}

\subsection{\spiderDeN: tracing the Cosmic Web using the density field}
\label{subsec:spider_density}
The density is one of the obvious fields used for environmental detection due to the sharp contrast of the clusters and filaments compared to most of remaining void dominated volume. To apply the \spider algorithm on the density one has to just insert the density $\delta$ in \eq{eq:hessian_fourier}. For simplicity, we denoted this method as \spiderDen. To better understand the behaviour of the \spiderDen method we need to rewrite the Hessian matrix (see \eqs{eq:hessian_fourier}{eq:hessian_kernel}) to:
\begin{equation}
     \hat{H}_{ij,R_n}(\Vector{k}) = -\frac{k_ik_j}{k^2} \hat\delta(\Vector{k})\;\;k^2 R_n^2 \;\; e^{-k^2R_n^2/2},
     \label{eq:density_hessian_fourier}
\end{equation}
where the initial equation was multiplied by the unit factor $\frac{1}{k^2}{k^2}$. It is immediately obvious that the Hessian is given by two distinct parts: the tidal field\footnote{The first part of \eq{eq:density_hessian_fourier} is the same as the Fourier transform of the tidal field given by \eq{eq:tidal_field_k} in \refsec{subsec:spider_tidal} up to the multiplication factor $4\pi G \bar{\rho}$. This factor has no effect on the final results since it only rescales the Hessian eigenvalues.} multiplied by a band-pass filter. The band pass filter is made of two distinct components: the $k^2 R_n^2$ high pass filter and the Gaussian $e^{-k^2R_n^2/2}$ low pass filter. Simple calculations show that the maximum of the band pass filter is at $k=\frac{\sqrt{2}}{R}$, while the shape of the filter is very similar to the one in \reffig{fig:hessian_kernel}. Therefore detecting the Cosmic Web structures in the density field is equivalent to identifying those structures in a band pass filtered tidal field.

When applying the \spiderDen formalism, an additional step has to be taken on top of those described in  \refsec{subsec:spider_description} and mask the density field when detecting filaments and walls. For filaments identification we need to set the density to $0$ in the cluster regions. In the absence of this mask, the cluster regions will give a large, unrealistic, filamentary signature\footnote{The filamentary signature of cluster regions will be large, even though the filamentary shape strength $\mathcal{I}_\rmn{filament}$ given by \eq{eq:shape_strength} is small in those regions. This is since the filamentary signature (see \eq{eq:response}) depends on $|\lambda_2|$ which has very large values in the cluster regions and will compensate for the small values of $\mathcal{I}_\rmn{filament}$. So this additional $|\lambda_2|$ factor that discriminates between signal and noise also introduces false detections. More generally, this false detection problem is important when the typical values of the tracer field for the different environments are orders of magnitude apart. It can be easily corrected by using the masking procedure described in the text.}. Similarly when identifying walls, we need to set the density to $0$ in both the cluster and filament regions.

\begin{figure}
    \centering
    \includegraphics[width=0.7\linewidth,angle=-90.0]{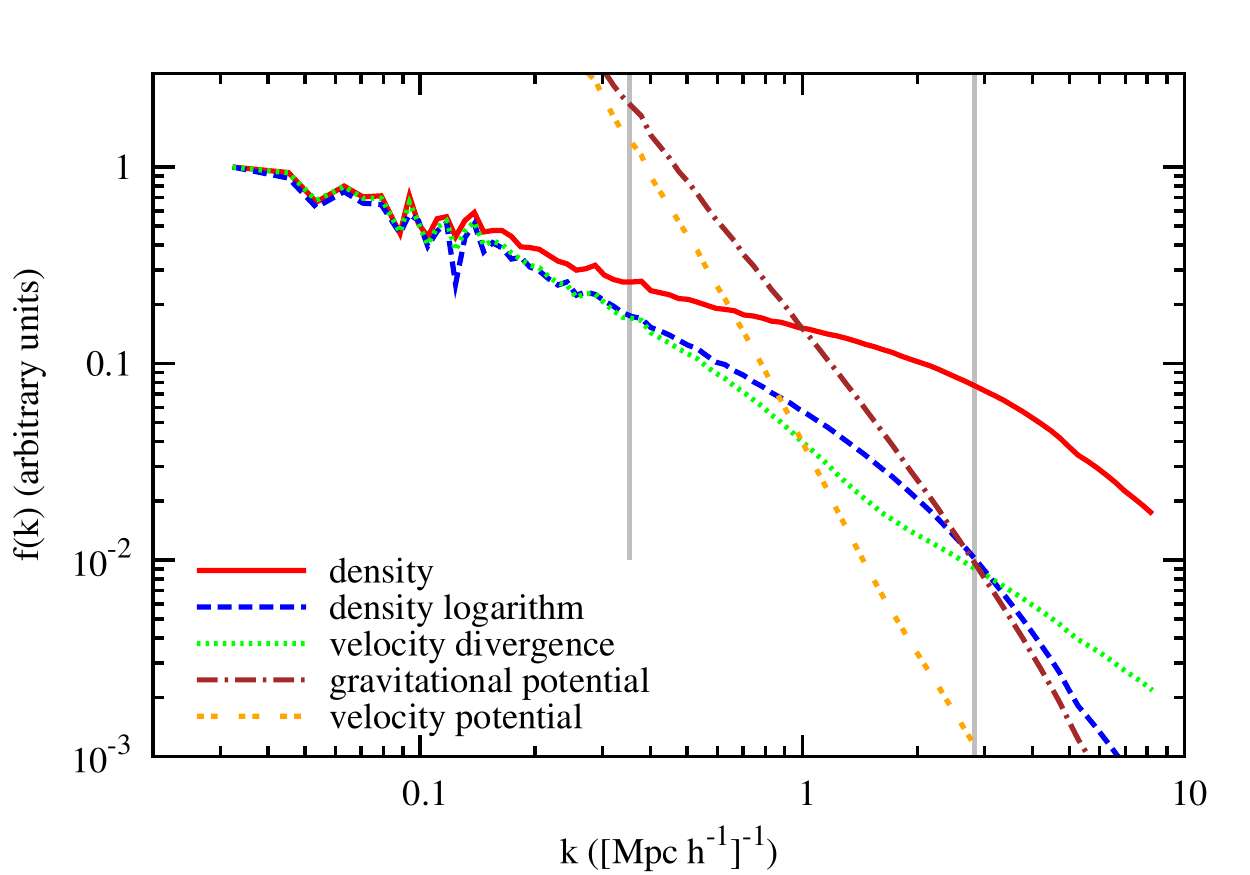}
    \caption{The Fourier transform amplitude for the density, density logarithm, gravitational potential, velocity divergence and velocity potential fields. \MC{The spectra were obtained directly from the DTFE interpolated fields used as input for the \spider method.} The curves were shifted vertically to better emphasize the differences between the two gray lines which mark the peak of the $\hat{\mathcal{H}}_R$ function with $R=4$ (left line) and $R=0.5\Mpch$ (right line). The two smoothing radii represent the upper and lower limits of the smoothing scales set used in the \spider algorithm.}
    \label{fig:input_field_spectra}
\end{figure}

\subsection{\spiderTidaL: tracing the Cosmic Web using the tidal field}
\label{subsec:spider_tidal}
The tidal field is the driver of anisotropic gravitational collapse and it is an essential ingredient for the formation and evolution of the Cosmic Web \citep{Zeldovich70,Gurbatov89,Bondweb96}. It is only natural to use it for the detection and understanding of the cosmic structures \citep{Hahn2007a,Forero-Romero09}. This mode of the \spider method is indicated as \spiderTidaL.

The tidal field is given by
\begin{equation}
    T_{ij}(\Vector{x}) = \frac{\partial^2 \phi_\rmn{grav}(\Vector{x}) } {\partial x_i\partial x_j},
    \label{eq:tidal_field}
\end{equation}
with $\phi_{grav}$ the gravitational potential. The latter is related to the density via the Poisson equation:
\begin{equation}
    \nabla^2\phi_{\rmn{grav}}(\Vector{x}) = 4\pi G \bar{\rho} \delta(\Vector{x}),
    \label{eq:gravitational_potential}
\end{equation}
where $G$ is the gravitational constant, $\bar{\rho}$ is the average matter density and $1+\delta(x)=\rho(\Vector{x})/\bar{\rho}$ is the overdensity. The Poisson equation is easily solved in Fourier space to obtain:
\begin{equation}
     \hat \phi_{\rmn{grav}}(\Vector{k}) = -4\pi G \bar{\rho} \frac{1}{k^2}  \hat\delta(\Vector{k}),
    \label{eq:gravitational_potential_k}
\end{equation}
leading to the following expression for the Fourier components of the tidal field:
\begin{equation}
     \hat T_{ij}(\Vector{k}) = 4\pi G \bar{\rho} \frac{k_ik_j}{k^2}  \hat\delta(\Vector{k}).
    \label{eq:tidal_field_k}
\end{equation}

Identifying the cosmic environments traced by the tidal field is done by applying the \spider algorithm on the gravitational potential. The Hessian matrix of the potential $\phi_{\rmn{grav}}$ given by \eq{eq:gravitational_potential_k} reduces to:
\begin{equation}
     \hat{H}_{ij,R_n}(\Vector{k}) = 4\pi G \bar{\rho} \frac{k_ik_j}{k^2} \hat\delta(\Vector{k}) \;\; e^{-k^2R_n^2/2}.
     \label{eq:tidal_hessian_fourier}
\end{equation}
This is exactly the tidal field smoothed over with a Gaussian filter. 

\begin{figure}
    \centering
    \includegraphics[width=\linewidth]{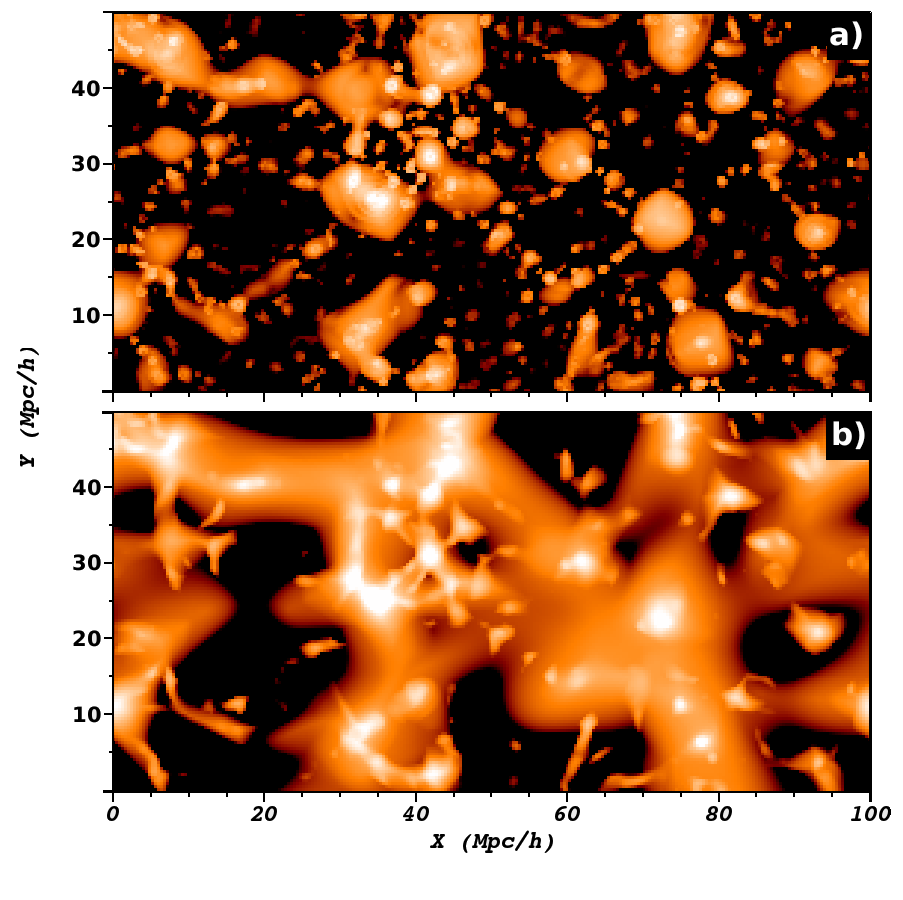}
    \caption{The filament signature computed using: a) \spiderDen and b) \spiderTidal methods. The graph shows a thin $1\Mpch$ slice. The white, orange and black correspond to high, medium and low signature values.}
    \label{fig:env_signature_comparison}
\end{figure}

The difference between the Hessian matrix of \spiderTidal and \spiderDen consists in the additional $k^2R_n^2$ high pass filter present in the case of the second method -- compare Eqs. \eqref{eq:density_hessian_fourier} and \eqref{eq:tidal_hessian_fourier}. Therefore variations in the result of the two methods comes from excluding the low frequency modes in the \spiderDen case and not in the \spiderTidal. The same conclusion can be reached by looking at the Fourier transform amplitude of the input fields: the density versus the gravitational potential. These are shown in \reffig{fig:input_field_spectra}. The first is more flat, while for the second the low frequencies have much larger amplitudes. This means that the large scale modes give a much larger contribution for \spiderTidal than for \spiderDeN. These effects are illustrated in \reffig{fig:env_signature_comparison}. The \spiderDen environments have a very clumpy appearance and are very sensitive to small scale structures. On the other hand, \spiderTidal is only responsive to the large scale modes and cannot trace the smaller details of the matter distribution.

It is important to note that the environment characteristics in the gravitational potential are different from the ones in the density field. According to the Cosmic Web theory, the clusters, filaments and walls are given by the strength and sign of the first, second and third eigenvalues of the tidal field \citep{Bondweb96}. This can be easily implemented within the \spider framework by changing the environmental signature from \eq{eq:response} to:
\begin{equation}
    \mathcal{S} = \theta(\lambda_{a}) \; \lambda_{a}
    \label{eq:response_tidal}
\end{equation}
with $a=1$ for clusters, $a=2$ for filaments and $a=3$ for walls.

As in the case of the \spiderDen method, we need to apply a cluster mask when identifying filaments and a combined cluster and filament mask when identifying walls. This procedure sets the density to $0$ in the mask regions, after which the gravitational potential is computed using \eq{eq:gravitational_potential}.

\subsection{\spiderDenloG: tracing the Cosmic Web using the density logarithm}
\label{subsec:spider_den_log}
Using the density logarithm for Cosmic Web detection is an approach that has not been explored until now. We were motivated to apply the \spider algorithm to the density logarithm and not the density itself because of a  multitude of reasons:
\begin{itemize}
    \item The \spider method works best when all structures correspond to similar values in the input field, while the density ranges from $0.01$ in underdense versus $10^4$ and higher in overdense regions. We expect to find structures over orders of magnitude in density values and simply using density biases the results towards high density structures. By taking the density logarithm the orders of magnitude difference is reduced to values of $-2$ in voids to around $4$ in cluster regions\footnote{Note that the DTFE density field will always have a density different from zero even in the emptiest voids. The typical DTFE density contrast in voids at redshift $z=0$ is $0.01$ to $0.1$.}.
    \item The non-linear density field is close to a lognormal distribution, when smoothed on scales of a few Mpc \citep{1991MNRAS.248....1C}.
    \item The large scale structure is best made visible when rendering the density logarithm and not the density itself.
\end{itemize}

The \spiderDenlog method consists in replacing the input field $f$ in the \spider algorithm with $\log_{10}(1+\delta)$. The main difference between this method and \spiderDen consists in the reduced contrast between underdense and overdense regions as well as a much steeper spectrum towards large scales for the density logarithm (see \reffig{fig:input_field_spectra}). \MC{Because of the reduced contrast between underdense and overdense regions when looking at the density logarithm, there is no need to apply the mask described for the \spiderDen method.}

While we were motivated by the same reasons as above to develop the \Spider algorithm, there is a large difference between the \Spider and \spiderDenlog methods as will be clearly visible in the results of \refsec{sec:environments}. In \spiderDenlog we identify the Cosmic Web using the logarithm of the density $\log_{10}(1+\delta)$, while for the \Spider method we trace the environments using the density field smoothed with the \logFilter filter.

\subsection{\spiderVeldiV: tracing the Cosmic Web using the velocity divergence}
\label{subsec:spider_vel_div}
The \spiderDeN, \spiderTidal and \spiderDenlog methods use only half of the phase space, the positional information, for identifying the elements of the Cosmic Web. It is interesting from both a theoretical and practical point of view to see how the remaining phase space can also be used to trace the large scale structure. The natural candidates for this are the velocity divergence and the velocity shear, due to the one-to-one connection between these quantities and the density and tidal field in the linear regime.

The velocity divergence is defined as:
\begin{equation}
    \theta(\Vector{x}) = \frac{1}{H}\Vector{\nabla}\cdot\Vector{v}(\Vector{x})
    \label{eq:velocity_divergence}
\end{equation}
where we divide by the Hubble factor $H$ such that $\theta$ is a unitless quantity. The velocity divergence is easily computed as an output of the DTFE method \citep{1996MNRAS.279..693B,2007MNRAS.382....2R}. According to linear theory, the velocity divergence is related to the density field via:
\begin{equation}
    \theta(\Vector{x}) = -f \delta(\Vector{x}),
    \label{eq:density_velocity_divergence}
\end{equation}
with $f$ the linear velocity growth factor \citep[see][]{Peebles80}. So in the linear regime, any structure in the density field should also be present in velocity divergence. A similar relation between $\theta$ and $\delta$ holds true also for the non-linear regime, but in a more complex way \citep[for details see][]{Nusser91,Chodorowski97,Bernardeau99}. The differences between the structures detected using density versus velocity divergence probe the effects of the non-linear evolution on the Cosmic Web components.

The \spiderVeldiv method is the application of the \spider algorithm on the negative of the velocity divergence $-\theta$. We choose the minus sign because of \eq{eq:density_velocity_divergence}. The \spiderVeldiv Hessian matrix is given by:
\begin{equation}
     \hat{H}_{ij,R_n}(\Vector{k}) = \frac{k_ik_j}{k^2} \hat\theta(\Vector{k})\;\;k^2 R_n^2 \;\;  e^{-k^2R_n^2/2}.
     \label{eq:vel_div_hessian_fourier}
\end{equation}
This is the product of the velocity shear given by \eq{eq:velocity_shear_divergence} multiplied by a band pass filter. This is exactly the same as for the \spiderDen method, but with the tidal field replaced by the velocity shear. The main difference between \spiderVeldiv and \spiderDen can be easily seen in \reffig{fig:input_field_spectra}: the velocity divergence and density have the same Fourier components at large scales, but the density has a more flattened drop at smaller scales.

\MC{In contrast to the \spiderDen method, we choose not to apply a mask for the velocity related methods. While there are still orders of magnitude variation in the velocity divergence between overdense and underdense regions, this difference is not as large as for the density field. The major challenge in applying a mask arises because the velocity divergence can take both positive and negative values, so there is no a priori well motivated value that we can use in the mask regions.}

\subsection{\spiderVelsheaR: tracing the Cosmic Web using the velocity shear}
\label{subsec:spider_vel_shear}
The velocity shear is the symmetric part of the velocity gradient, with the $ij$ component defined as:
\begin{equation}
    \sigma_{ij}(\Vector{x}) = \frac{1}{2H} \left( \frac{\partial v_j}{\partial x_i} + \frac{\partial v_i}{\partial x_i}\right)
    \label{eq:velocity_shear}
\end{equation}
where $v_i$ is the $i$ component of the velocity. We normalize the velocity shear by the Hubble constant to keep the same notations as in the case of the velocity divergence. To obtain the velocity shear, the velocity is rewritten as the sum of the potential and rotational flows:
\begin{equation}
    \Vector{v} =\nabla \phi_\rmn{vel} + \nabla\times \Vector{A_\rmn{vel}}
    \label{eq:velocity_components}
\end{equation}
where $\phi_\rmn{vel}$ is the scalar velocity potential and $A_\rmn{vel}$ is the vector potential. Inserting this last equation into the velocity shear gives:
\begin{equation}
    \sigma_{ij}(\Vector{x}) = \frac{1}{H} \frac{\partial^2 \phi_\rmn{vel}(\Vector{x})}{\partial x_i\partial x_j}.
    \label{eq:velocity_shear_potential}
\end{equation}
Therefore, the velocity shear depends only on the velocity potential. It is interesting to notice that the velocity divergence is also given only by the velocity potential via:
\begin{equation}
    \theta = \frac{1}{H} \nabla^2 \phi_\rmn{vel}.
    \label{eq:velocity_divergence_potential}
\end{equation}
This last equation can be inverted and used to solve for the potential $\phi_\rmn{vel}$, to obtain its Fourier components as:
\begin{equation}
    \hat\phi_\rmn{vel}(\Vector{k}) = -H\frac{1}{k^2} \hat\theta (\Vector{k}).
    \label{eq:velocity_potential_k_space}
\end{equation}
On the basis of this equation, one can infer the velocity potential starting from the velocity divergence output of the DTFE method. Combining these, we obtain that the components of the velocity shear can be expressed in terms of $\theta$ as:
\begin{equation}
    \hat\sigma_{ij}(\Vector{k}) = \frac{k_ik_j}{k^2} \hat\theta(\Vector{k}).
    \label{eq:velocity_shear_divergence}
\end{equation}
Thus the relation between the velocity shear and the tidal tensor in the linear regime is given by:
\begin{equation}
    \sigma_{ij}=-fT_{ij}.
    \label{eq:velocity_shear_tidal}
\end{equation}

\begin{figure}
    \centering
    \includegraphics[width=\linewidth]{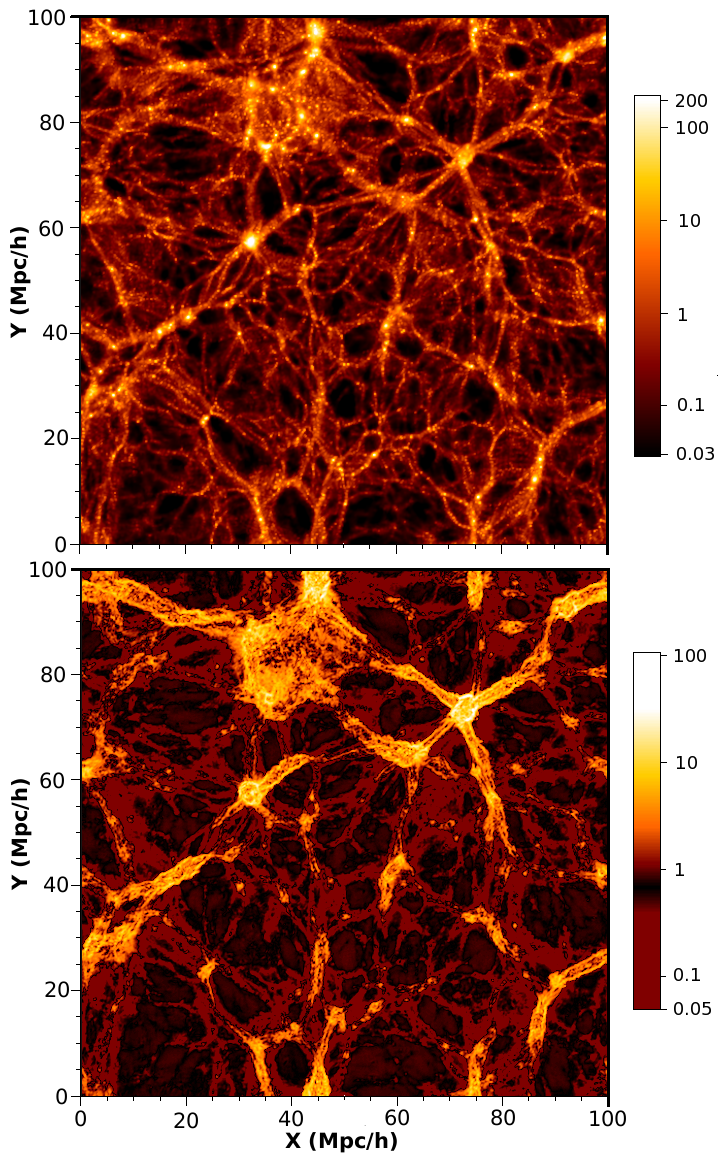}
    \caption{A $1\Mpch$ slice from the N-body simulation illustrating the DTFE density $1+\delta$ (upper panel) and absolute value of the velocity divergence $\theta$ (lower panel).}
    \label{fig:density_divergence_slice}
\end{figure}
\begin{figure}
    \centering
    \includegraphics[width=\linewidth]{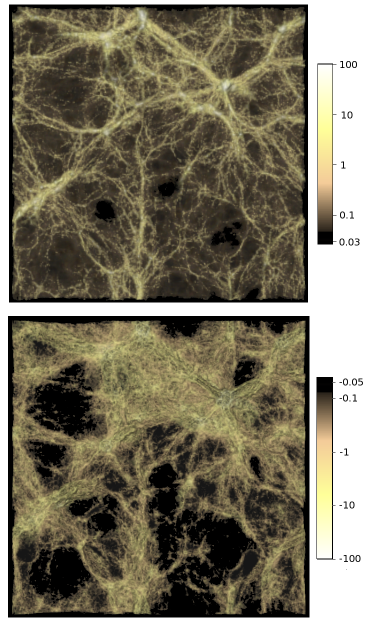}
    \caption{A 3D volume rendering of the density $1+\delta$ (left panel) and velocity divergence $\theta$ (right panel). Note that we only show the negative values of the velocity divergence. The picture represents a $100\times100\times10(\Mpch)^3$ volume in an N-body simulation. We used the same volume to illustrate the Cosmic Web environments in Figures \ref{fig:env_filaments}-\ref{fig:env_halos}.}
    \label{fig:density_divergence_box}
\end{figure}

The \spiderVelshear method involves using the negative of the velocity potential $-\phi_\rmn{vel}$ as input field to the \spider algorithm. The negative sign comes, as in the case of \spiderVeldiV, from the minus in relation \eq{eq:velocity_shear_tidal}. Inserting the expression for $\phi_\rmn{vel}$ in the Hessian matrix gives:
\begin{equation}
     \hat{H}_{ij,R_n}(\Vector{k}) = -\frac{k_ik_j}{k^2} \hat\theta(\Vector{k}) \;\;  e^{-k^2R_n^2/2}.
     \label{eq:vel_shear_hessian_fourier}
\end{equation}
As in the case of the \spiderTidal method, the environment characteristics are different in velocity shear compared to velocity divergence. So the environment signature has to be changed to the expression given by \eq{eq:response_tidal}.

\section{N-body simulations and halo catalogues}
\label{sec:nbody}

To test our structure finding algorithms, we apply them to cosmological N-body simulations containing only dark matter particles. We adopted the \lcdm cosmological model with $\Omega_m=0.26$, $\Omega_\Lambda=0.74$, $h=0.71$, $\sigma_8=0.8$ and $n_s=1$. We performed two $512^3$ particle simulations in a $100\Mpch$ and $200\Mpch$ periodic boxes. The force resolution was fixed in comoving coordinates up to $z=4$, to $15\kpch$ and $21\kpch$. Afterwards it was fixed in physical coordinates to $5\kpch$ and $7\kpch$, respectively. 

The $100\Mpch$ simulation (mass resolution of $5.4\times 10^{8}\Msolar$) was chosen in order to resolve haloes to a few times $10^{10} \Msolar$ and at the same time to have a reasonable cosmological volume whose smallest mode is still evolving linearly. We used this small volume simulation for visualization and resolution studies. The $200\Mpch$ simulation (mass resolution of $4.3\times 10^{9}\Msolar$) is used for computing the quantitative results since a larger volume gives better statistics.

The simulations were performed using the public version of the parallel Tree-PM code Gadget2 \citep{Springel2005a} on a Linux cluster at the University of Groningen, Netherlands. The initial conditions for both simulations were generated at the $z=50$ redshift using the transfer function given by \citet{1986ApJ...304...15B}.

\subsection{Density and velocity divergence fields}
\label{subsec:nbody_density}
The output of the N-body simulation consists of a discrete set of particles. This needs to be interpolated to a continuum volume-filling density and velocity divergence fields that will be used as input for the Cosmic Web environment detection algorithm. It is crucial for the environment detection procedure, especially for anisotropic features such as filaments and walls, that the interpolation method used to obtain the continuous fields retains all the scale and geometry information of the discrete galaxy or particle distribution.

For these reasons we use the Delaunay Tessellation Field Estimator (DTFE), introduced by \citet{2000A&A...363L..29S} \citep[for additional details see][]{2009LNP...665..291V,Cautun2011}, to reconstruct the underlying density and velocity divergence fields. 
For the environment detection algorithm, the DTFE method has the following important advantages:
\begin{enumerate}
    \item[$\bullet$] Preserves the multi-scale character of the discrete distribution.
    \item[$\bullet$] Preserves the local geometry of the discrete distribution.
    \item[$\bullet$] Does not depend on user defined parameters or choices.
\end{enumerate}

The continuous DTFE density and velocity divergence fields are sampled on a $256^3$ and $512^3$ grid for the $100\Mpch$ and $200\Mpch$ simulations respectively, such that there is a $0.4\Mpch$ grid spacing in both cases. \reffig{fig:density_divergence_slice} shows a thin slice of the grid sampled density and velocity divergence fields. Note the level of detail in the structures, even inside voids, and the one-to-one correspondence between the DTFE density and velocity divergence features. \reffig{fig:density_divergence_box} gives a 3D rendering of the density and velocity divergence fields in a larger volume - the same volume that later on will be used to visualize the Cosmic Web environments.

\subsection{Halo and subhalo catalogues}
\label{subsec:nbody_halo_catalogues}
We use the AMIGAs Halo Finder (AHF) by \citet{Knollmann2009} to identify the dark matter haloes. The AHF halo finder is the successor of the MHF halo finder by \citet{2004MNRAS.351..399G}. AHF uses adaptive mesh refinement to identify the density peaks which it classifies as the halo and subhalo centres. Afterwards it grows the objects around their centres until the spherically averaged density contrast reaches the virial density\footnote{The virial density is automatically computed by AHF and depends on both cosmological parameters and redshift.}. The last step consist in removing the gravitationally unbound particles. The AHF halo and subhalo catalogues are complete up to haloes with $50$ or more particles \citep[for a complete description see][]{Knollmann2009}.

\section{The Cosmic Web Environments}
\label{sec:environments}

\subsection{Clusters}
\label{subsec:clusters}
The point-like objects detected by the \spider and \Spider methods correspond to large overdensities in the density field. They range from very massive to very small mass objects. We will see later in this section that these objects correspond to dark matter haloes. On the other hand, the cosmic clusters are the largest and most recent to form virialized objects \citep{2005RvMP...77..207V}. To be able to identify our point-like objects with actual clusters, we need to limit our detections to only the most massive objects. In this study we consider as clusters the objects with mass larger than $5\cdot10^{13}\Msolar$. This is a compromise between studying the most massive objects and having a large sample of such objects in our simulation. 

\begin{figure}
    \centering
    \includegraphics[width=0.7\linewidth,angle=-90.0]{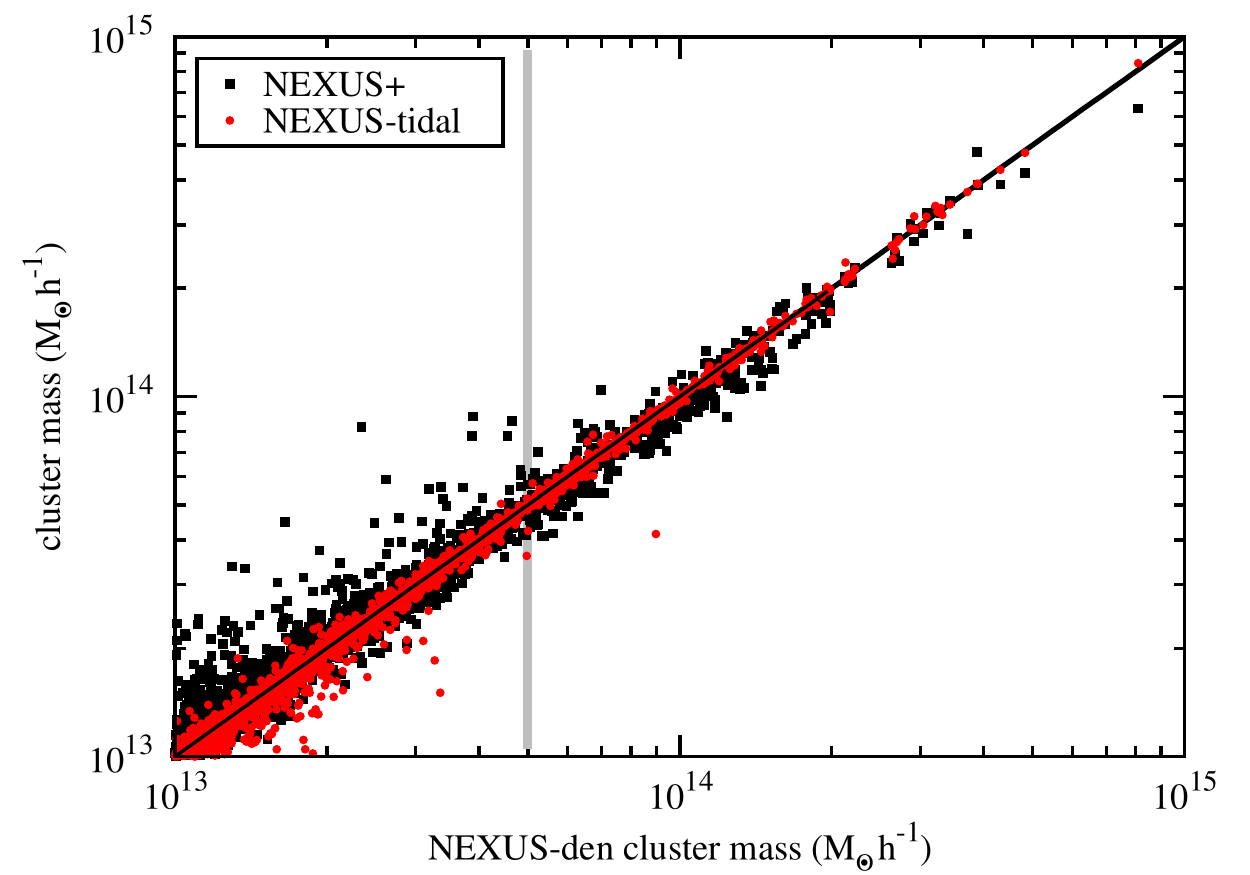}
    \caption{Comparison between the mass of cluster-like objects detected using the \spiderDeN, \spiderTidal and \Spider methods. The gray vertical line delineates the objects with mass larger than $5\cdot10^{13}\Msolar$ \MC{while the diagonal black line shows a one-to-one relationship}.}
    \label{fig:node_comparison}
\end{figure}

\begin{figure}
    \centering
    \includegraphics[width=0.69\linewidth,angle=-90.0]{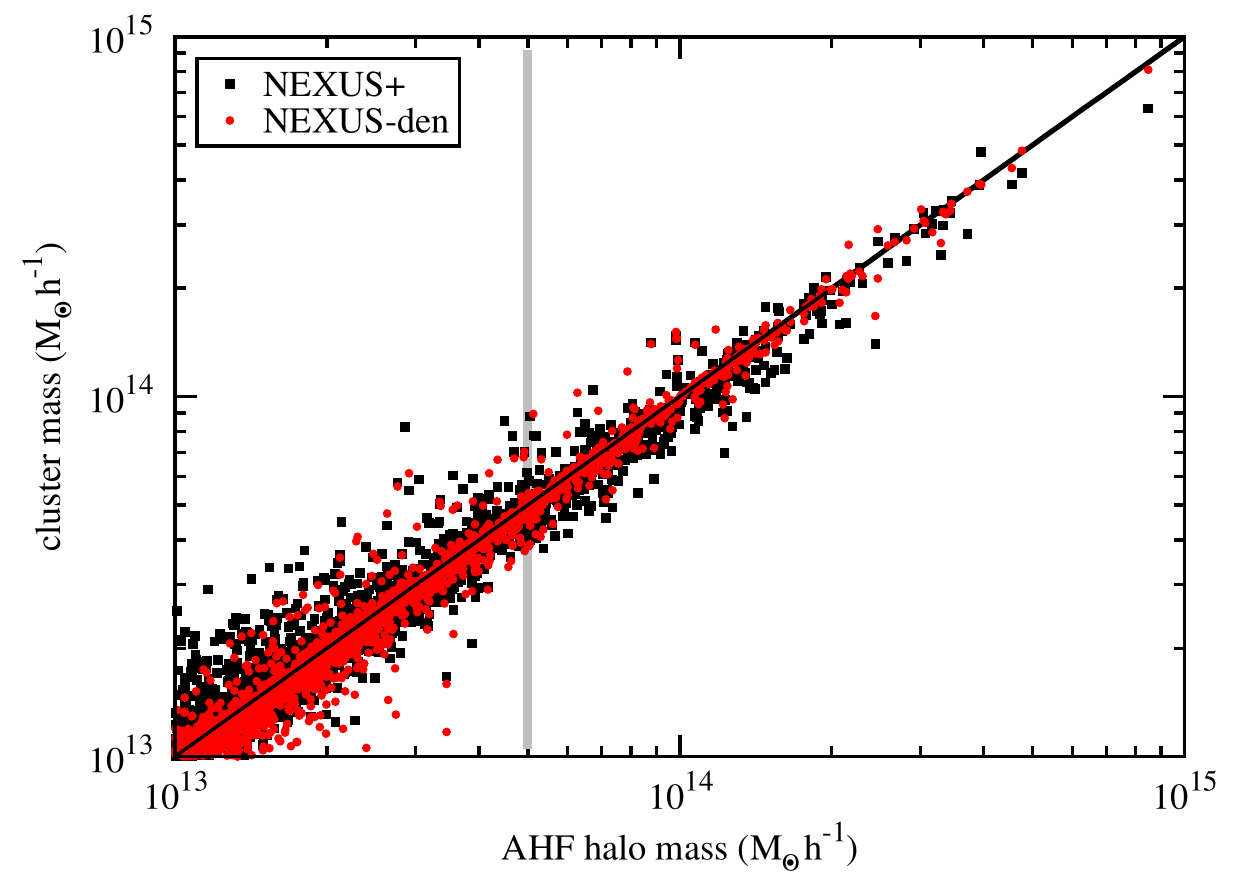} \\
    \includegraphics[width=0.69\linewidth,angle=-90.0]{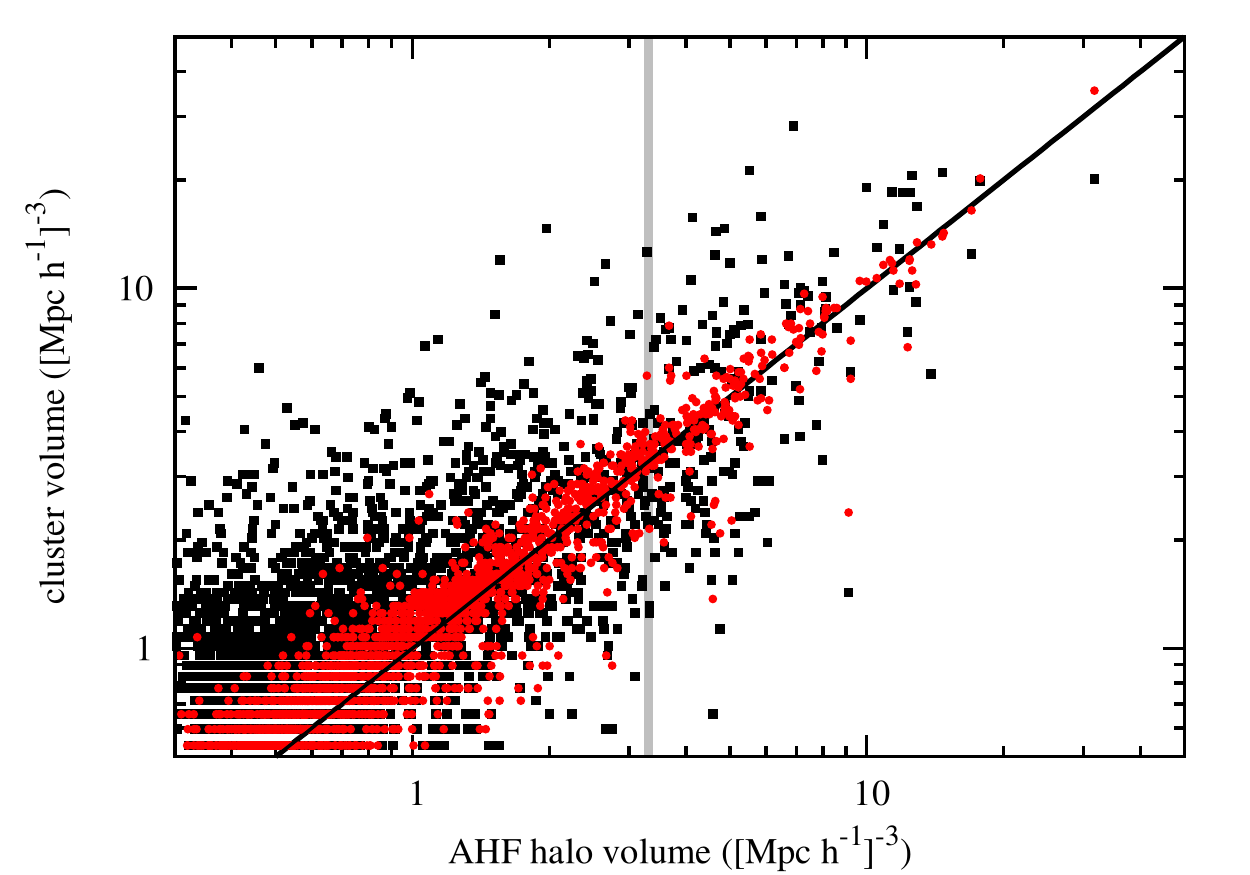} \\
    \includegraphics[width=0.69\linewidth,angle=-90.0]{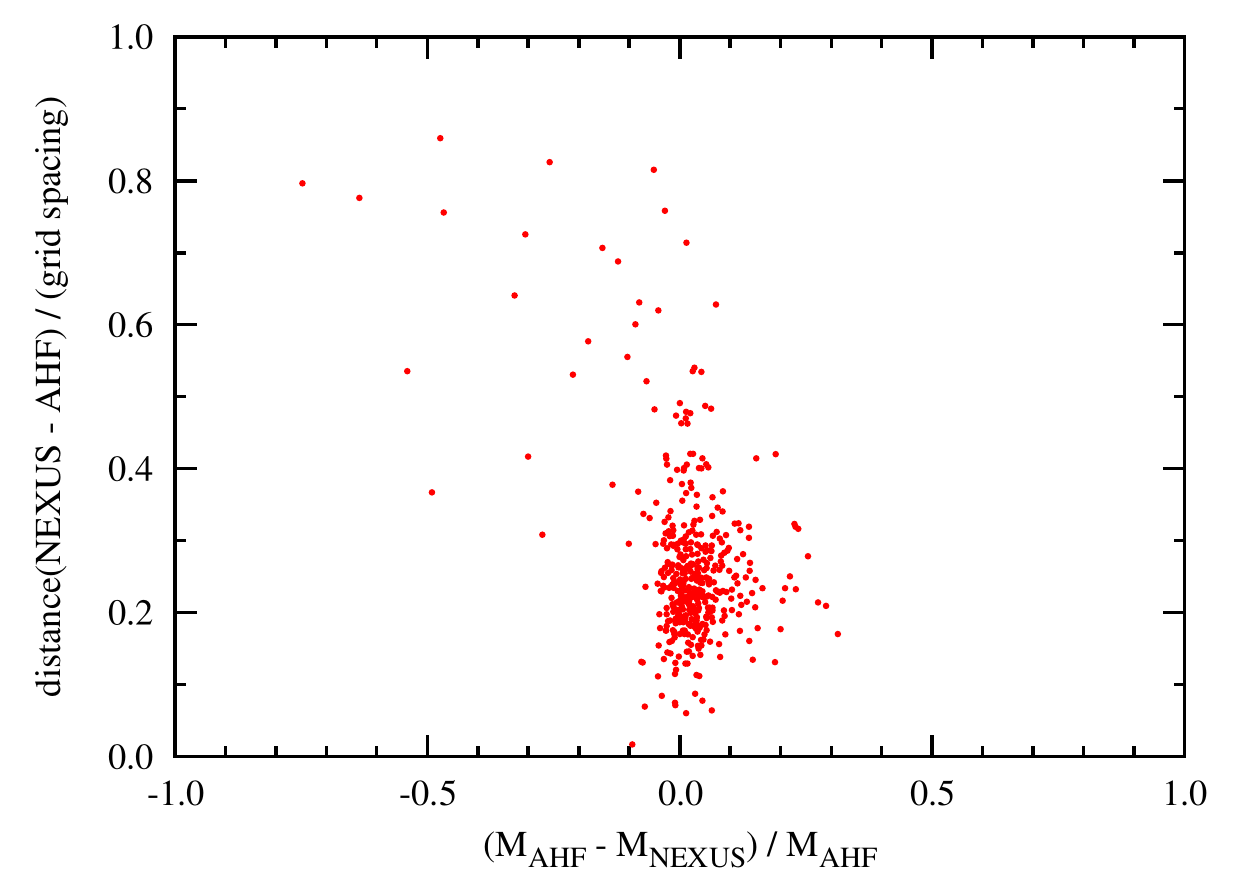}
    \caption{Comparison of cluster-like objects with their corresponding AHF haloes. The top panel shows the mass comparison while the middle panel shows the volume comparison for the \Spider and \spiderDen objects. The gray vertical line delineates the objects with mass larger than $5\cdot10^{13}\Msolar$ \MC{while the diagonal black line shows a one-to-one relationship}. The lower panel gives the mismatch in center position and mass difference between the AHF haloes and \spiderDen clusters (only for objects with mass larger than $5\cdot10^{13}\Msolar$). }
    \label{fig:AHF_node_comparison}
\end{figure}

The cluster detection method performs very well for the \spiderDeN, \spiderTidal and \Spider methods. In contrast, it does not give reliable detections for the \spiderVeldiv and \spiderVelshear methods. We suspect this deficiency is due to high vorticity in the cluster regions, vorticity that is not captured in the velocity divergence or shear fields and therefore is not taken into account in our methods.

The \spiderDeN, \spiderTidal and \Spider methods detect the same clusters, with similar mass and volume associated to each object. This can clearly be seen in the mass comparison plot shown in \reffig{fig:node_comparison}. The \spiderDen and \spiderTidal methods give very similar results, while we find a larger scatter when comparing these results with \SpideR. 

There is a very close connection between the cluster-like objects of our methods and the most massive dark matter haloes. This is shown in \reffig{fig:AHF_node_comparison} where we use AHF to identify the dark matter haloes. The top panel shows that there is a very good correlation between the AHF halo mass and its corresponding object identified using our methods. We show that this relation holds down to masses of $10^{13}\Msolar$. This lower limit is a limitation of the grid spacing size and not of the method. We see that \spiderDeN, and also \spiderTidal (not shown), give a much better correlation of the cluster mass with the corresponding AHF halo mass, while in the case of \Spider there is a larger scatter. A similar comparison is done between the cluster volumes and the AHF halo volume - see center frame of \reffig{fig:AHF_node_comparison}. Again there is a good match between the methods, but with a much wider scatter than in the case of the mass comparison. This is understandable since the volume is much more sensitive to the lower density regions around clusters, while the mass is dominated by the inner high density regions. In this case the scatter is especially large in the \Spider results, than for the \spiderDen and \spiderTidal methods.

Finally we compare the \spiderDen results and AHF haloes only for the most massive objects. In the bottom panel of \reffig{fig:AHF_node_comparison} it can be seen that the clusters agree within $10\%$ to their corresponding AHF halo mass and that their centres are at most $0.4$ grid spacing distance from the AHF halo center. The few outliers are objects that are merged in one of the methods and detected as distinct objects in the second one. 

We see that \Spider is less reliable in the detection of Cosmic Web clusters than the \spiderDen or \spiderTidal methods. The very localized nature of the \logFilter filter means that the contribution of the highly dense center does not have a large effect on the periphery of the cluster. Because of this the outer boundaries of the \Spider clusters are more dependent on the substructure at the periphery. One can overcome this "weakness" in the \Spider method by identifying the clusters using \spiderDen or \spiderTidaL.

\subsection{Filaments}
\label{subsec:filaments}

\begin{figure*}
    \centering
    \includegraphics[width=0.82\linewidth]{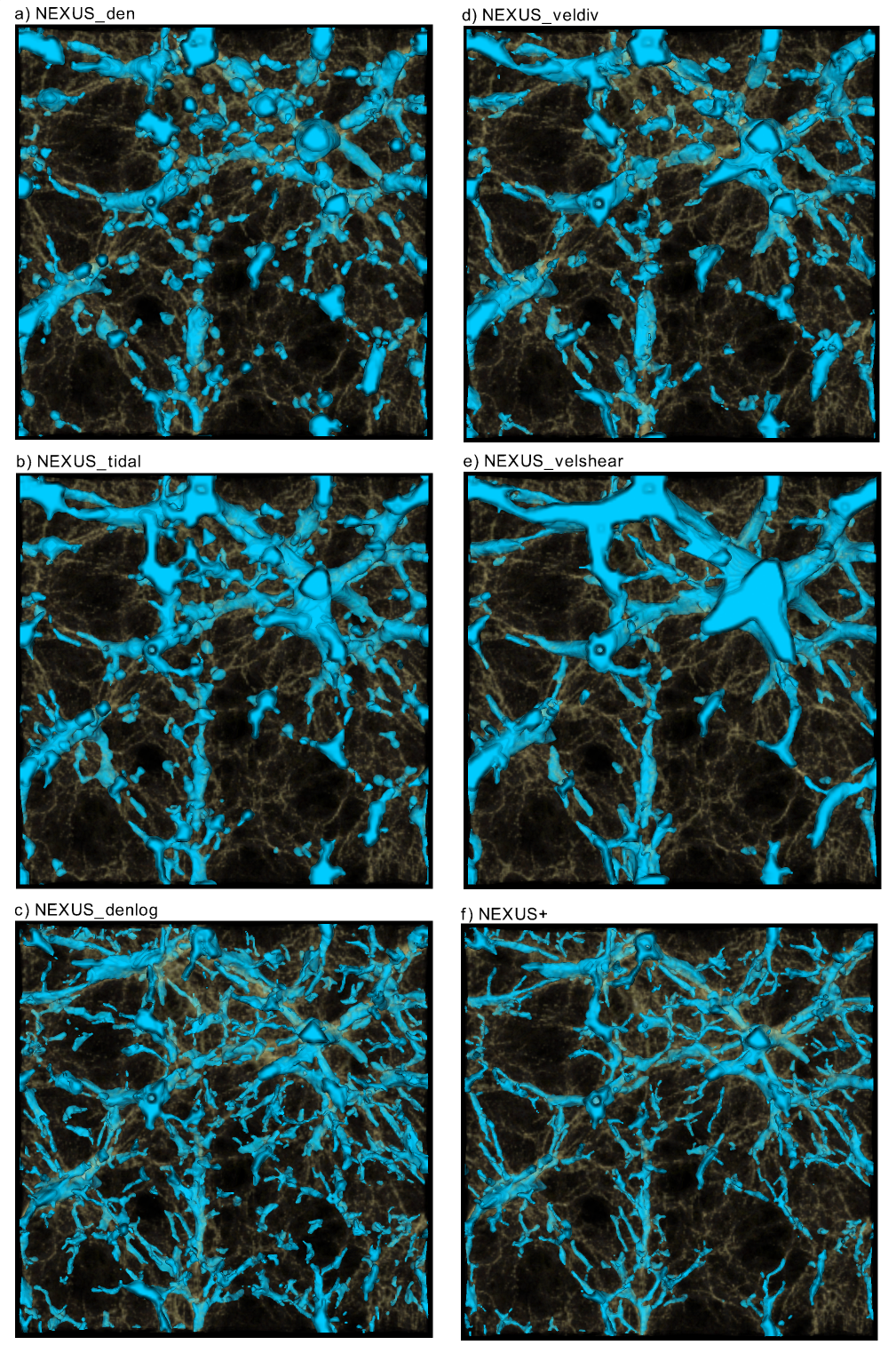}
    \caption{A 3D rendering of the filaments in a $100\times100\times10\MpchVolume$ volume of the N-body simulation. The faint background shows the density field. The filaments were obtained using: a) \spiderDeN, b) \spiderTidaL, c) \spiderDenloG, d) \spiderVeldiV, e) \spiderVelshear and f) \SpideR. This is the same region of the simulation volume as the density field shown in \reffig{fig:density_divergence_box}.}
    \label{fig:env_filaments}
\end{figure*}

After clusters, filaments are the most noticeable feature of the Cosmic Web. These structures are shown in \reffig{fig:env_filaments}, as identified by the six methods proposed in this work. To obtain these results we have restricted our discussion to the most significant filaments, which we define as any continuous region with a volume larger than $10\MpchVolume$. By doing so we discard small regions, typically around isolated haloes which even though show a local filament signature, are not embedded in the larger network.

Comparing the filamentary maps with the density field rendering of the same volume from \reffig{fig:density_divergence_box}, we see that all the methods succeed in identifying the strongest filaments, but there are differences when it comes to the smaller, less pronounced filaments. We immediately observe that the Cosmic Web filaments form an interconnected network, with the most massive filaments acting as the backbone. These very pronounced filaments branch into thinner ones that slowly disappear into lower density regions. The backbone filaments are clearly visible in the central panel while the branching into fainter structures is most pronounced in the lower-right panel of \reffig{fig:env_filaments}. A visual inspection of clusters (not shown here) shows that these reside at the intersection of the most prominent filaments, which serve as the highways along which matter is transported to the clusters.

A visual comparison of the filamentary maps obtained using the six methods leads to the following conclusions:
\begin{itemize}
    \item The filamentary structures in the \spiderDen and \spiderVeldiv are more clumpy and have less large scale cohesion when compared to the \spiderTidal and \spiderVelshear results. This is in line with our expectations, since the large scale modes bring a larger contribution for the latter methods (see \reffig{fig:input_field_spectra}).
    \item The methods using the tidal and velocity shear fields are biased towards the most significant structures and miss most of the filaments present in the less dense regions. This less pronounced filaments, marginally seen by the \spiderDen and \spiderVeldiv methods, are very well reproduced in the \spiderDenlog and \Spider methods. This supports the view that the Cosmic Web has structures present over a large range in density values.
    \item While the most pronounced filaments are detected by all the methods, their thickness varies between the different methods. The \spiderDenlog and \Spider filaments have typical diameters around $2\Mpch$, while for the rest the typical diameter is around $4\Mpch$. The thinner filaments mostly constitute the inner regions of the much thicker filaments detected using the other methods.
    \item While \spiderDenlog finds the same structures as the other methods in the high density regions, it finds a much richer filamentary network in the underdense regions. This is because any small changes in the density field in these regions can lead to a large contrast in the density logarithm. The Poissonian sampling noise is especially important for density determination in the underdense regions due to sparse sampling. This makes the \spiderDenlog method especially sensitive to Poissonian noise in the void-like regions.
\end{itemize}

To summarize, the variation in the six methods manifests itself as mostly differences in the detection of smaller filaments and thickness differences in the very prominent structures. Since the large scale modes contribute much more to \spiderTidal and \spiderVelsheaR, these methods identify only the largest filaments that correspond to the peaks of the large scale modes. On the other hand, the \spiderDenlog and \Spider methods are much more sensitive to the less pronounced structures, finding an important filamentary network also in the underdense regions. While \spiderDen and \spiderVeldiv are in between the two classes of results, they are much closer in character to the \spiderTidal and \spiderVelshear methods.

\subsection{Walls}
\label{subsec:walls}

\begin{figure}
    \centering
    \includegraphics[width=.85\linewidth]{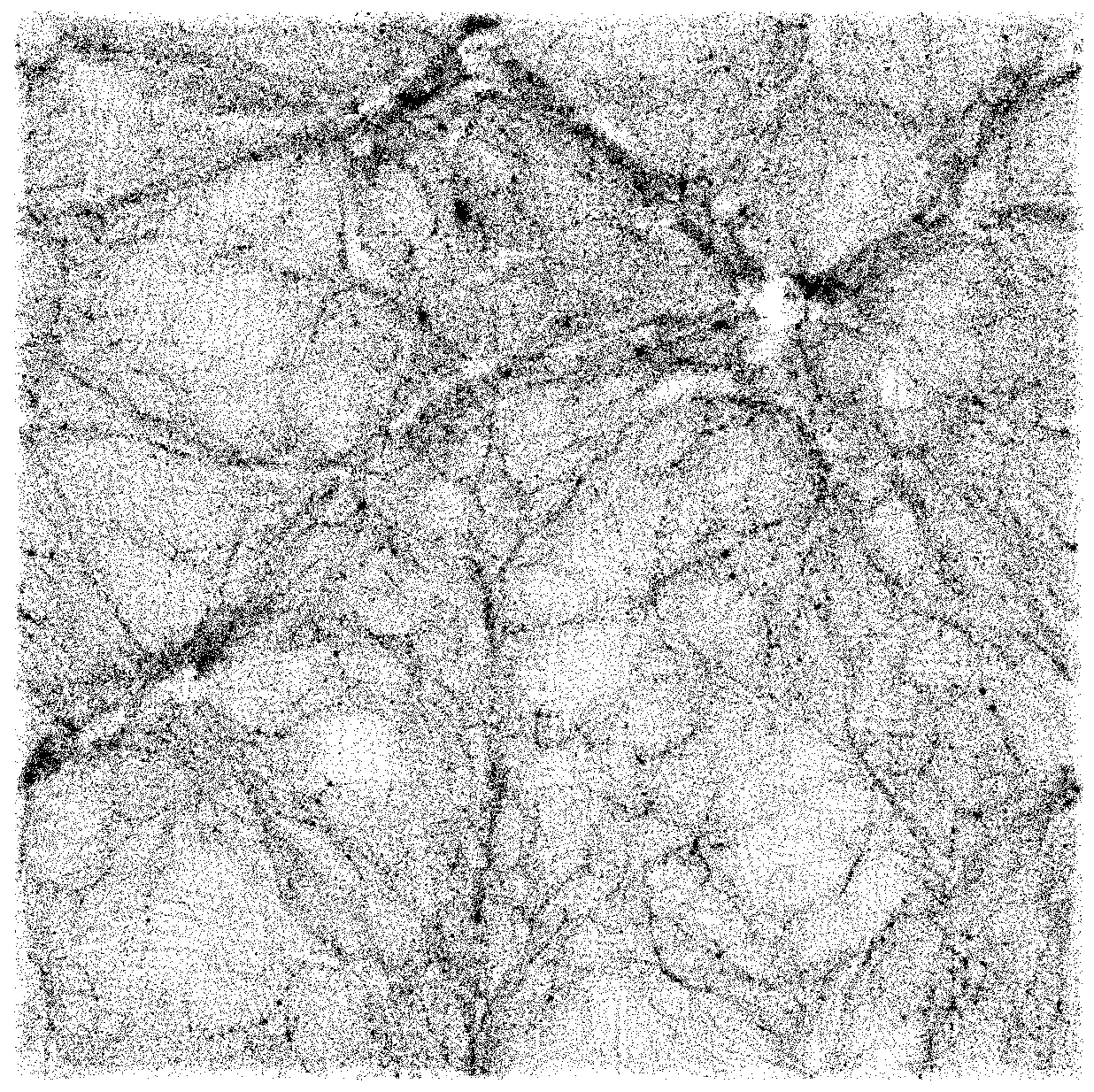}
    \caption{The dark matter particles left after taking out the particles located in cluster and filament environments. This is a projection of the volume of the N-body simulation as in Figures \ref{fig:density_divergence_box}, \ref{fig:env_filaments} and \ref{fig:env_walls}. }
    \label{fig:env_particles_walls}
\end{figure}

\begin{figure*}
    \centering
    \includegraphics[width=0.82\linewidth]{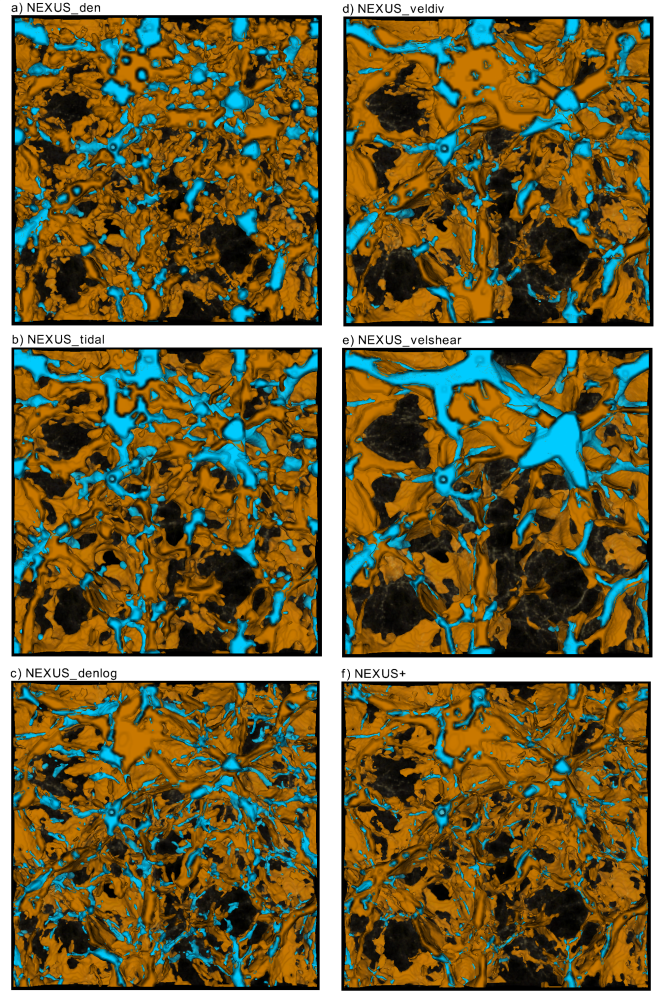}
    \caption{A 3D rendering of the walls (orange) in a $100\times100\times10\MpchVolume$ volume of the N-body simulation. The light blue depicts the filaments while the faint background shows the density field. The walls were obtained using: a) \spiderDeN, b) \spiderTidaL, c) \spiderDenloG, d) \spiderVeldiV, e) \spiderVelshear and f) \SpideR. This is the same region of the simulation volume as the density field shown in \reffig{fig:density_divergence_box} and the filaments in \reffig{fig:env_filaments}.}
    \label{fig:env_walls}
\end{figure*}

\begin{figure*}
    \centering
    \includegraphics[width=.85\linewidth]{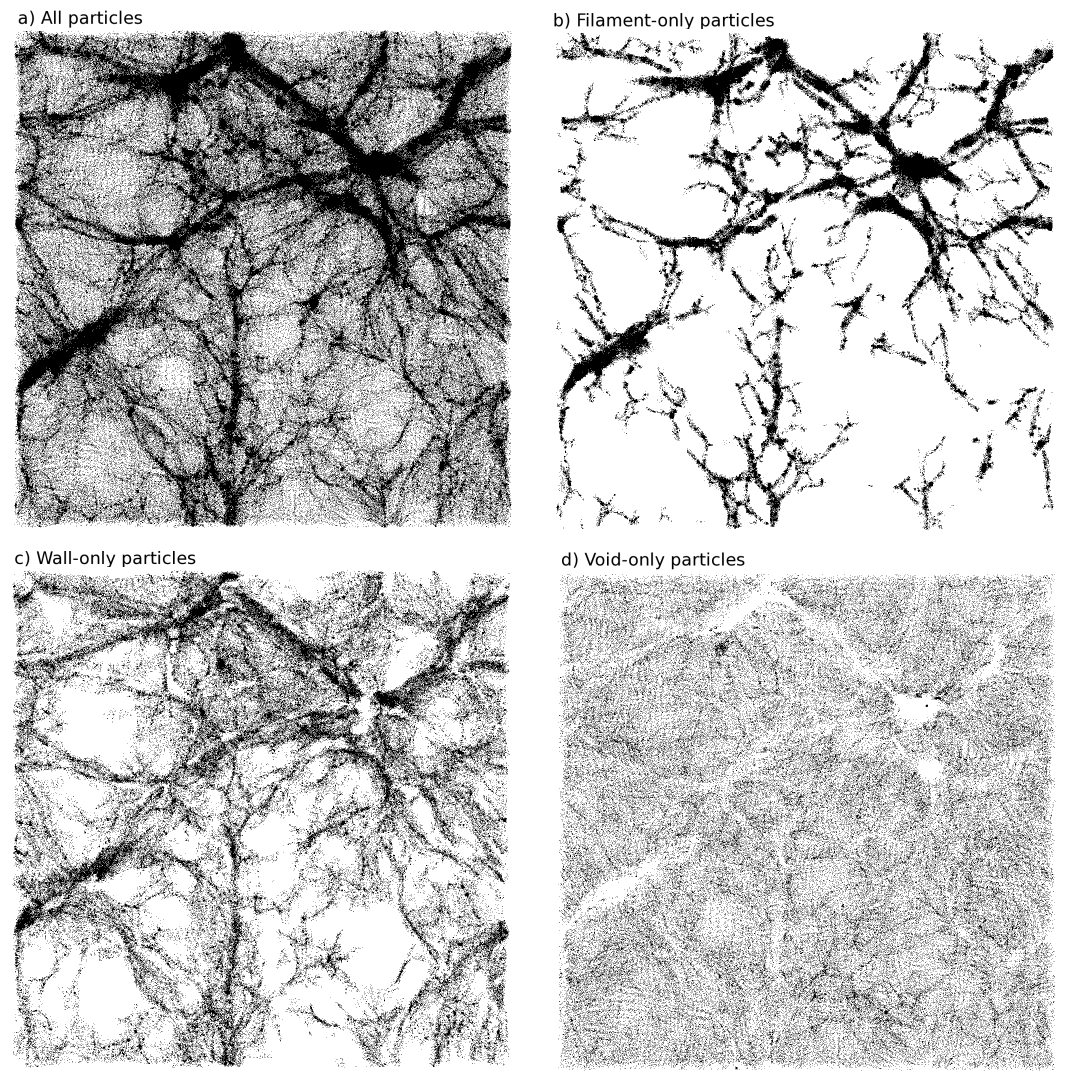}
    \caption{The dark matter particles in the different environments of the Cosmic Web. The panels gives: a) all, b) filament-only, c) wall-only and d) void-only particles. The environments where identified using the \Spider method.}
    \label{fig:env_particles}
\end{figure*}
\begin{figure*}
    \centering
    \includegraphics[width=.9\linewidth]{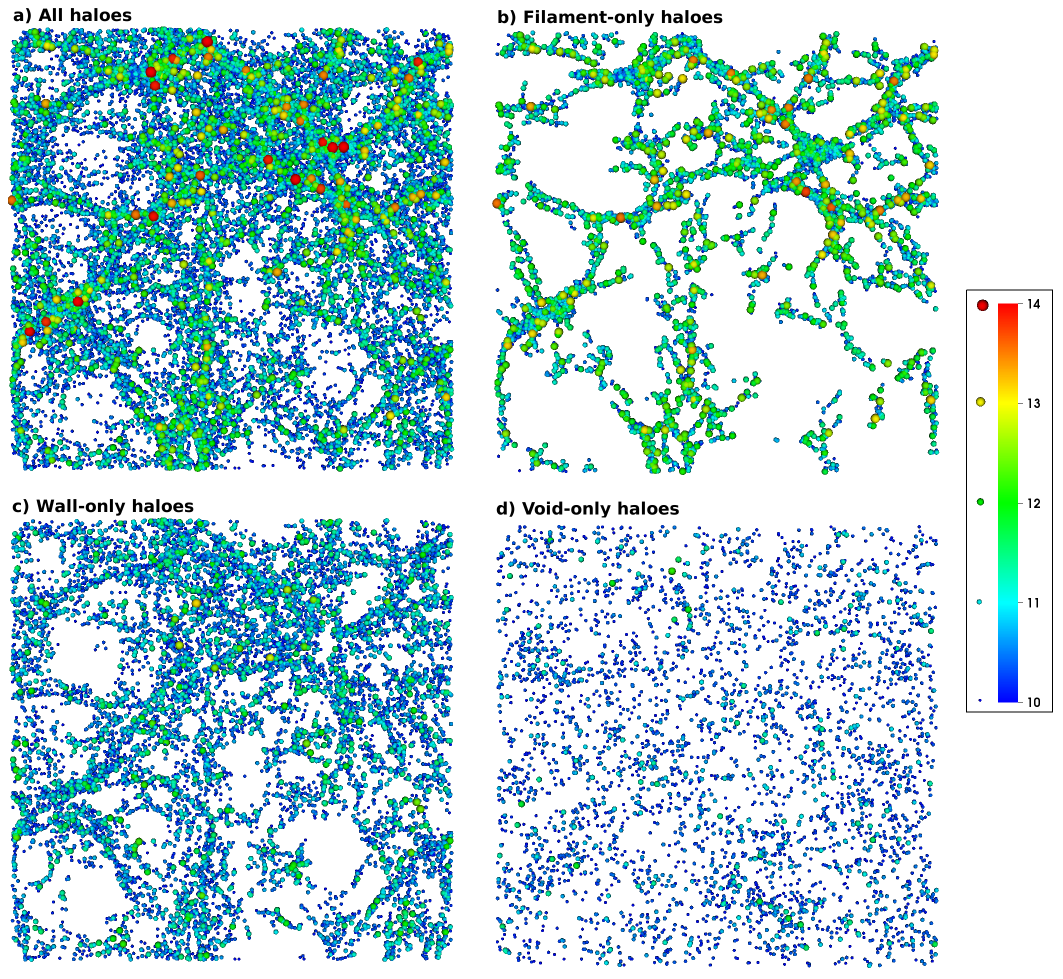}
    \caption{The dark matter haloes as a function of the environment they reside in. The panels gives: a) all, b) filament-only, c) wall-only and d) void-only haloes. The environments where identified using the \Spider method. The colour and size of the points is proportional to the mass of the halo they represent. The labels in the legend correspond to $\log_{10}(M/\Msolar)$. \MC{Note that most of the smaller mass haloes around massive ones are not visible due to the larger size with which we show the massive haloes.} }
    \label{fig:env_halos}
\end{figure*}

The wall environments detected using the six methods are presented in \reffig{fig:env_walls} where, on top of the walls in orange, we superimposed the filaments in light blue. In the case of filament identification one can use the density and velocity divergence maps to judge the success of the detection method, but this is much more difficult for walls. This is a consequence of both the smaller contrast and the planar nature of these structures. The presence of sheet-like structures in the distribution of matter on cosmological scales can be easily inferred from \reffig{fig:env_particles_walls}, where we show the dark matter particles after removing the particles located in clusters and filaments. The most striking structures are the line-like arrangements of particles visible especially in the upper part of the figure. These are sheets that are perpendicular on the projection plane and not filaments missed by our detection algorithm. There are additional walls along the projection plane (e.g. center upper part, between the three line-like structures) but these are less easily detected visually.

By comparing the particle distribution from \reffig{fig:env_particles_walls} to the \spider walls in \reffig{fig:env_walls} it is clear that the algorithm is very successful in identifying both the prominent as well as the tenuous Cosmic Web walls. The resulting objects form large continuous planar structures which delineate the different voids. The aspect of the sheets resembles very much that of walls of biological cells that have been stacked on top of each other, with the cells having a wide range of sizes. Even though the Cosmic Web walls are continuous for large portions of their surface, they are still punctured by a large number of holes that allow for void percolation. While these holes are dependent on the wall identification method, they are a sign of the diffuse and sparse nature of the cosmic walls.

As in the case of the filamentary network, the prominent walls are detected by all the methods, but there are differences when it comes to the sheets in the more underdense regions. Some of the most important features and differences between the six results can be summarized as:
\begin{itemize}
    \item The \spiderDen walls have a very clumpy appearance and this is also true to a lesser extent for the \spiderTidal results. This is in contrast with the other methods where the walls have a much more planar and sheet-like look. The clumpy appearance is due to the composition of walls, which are tenuous structures with sporadic haloes from place to place. The concentration of mass in the haloes compared to their neighbourhood regions gives the clumpy structure of the \spiderDen and \spiderTidal walls.
    \item The \spiderVeldiv and \spiderVelshear walls have a smooth planar appearance since the velocity field, compared to the density, is less affected by the sparse Poissonian sampling of the mildly underdense regions. These are the regions that make most of the volume in walls. We find the same smooth planar look also for the \spiderDenlog and \Spider walls.
    \item We find that for all the methods the larger gaps in the walls are present at the same locations, indicating that, at least occasionally, there is no clear boundary between adjacent voids.
    \item The \spiderVelshear method seems to be the most conservative tracer of walls. It detects only the most prominent wall regions.
\end{itemize}

A visual inspection favours the \SpideR, \spiderDenlog and \spiderVeldiv as best tracers of the cosmic sheets with the remaining methods either giving clumpy detections or missing some of the more tenuous structures. 

\MC{By comparing the filaments and walls in \reffig{fig:env_particles_walls} we find that most if not all the filament volume elements are embedded in walls. Though not shown, we find that also the clusters are fully embedded in filaments. This suggests that the strict classification where each volume element is assigned to a single environment can be extended by realizing that clusters are embedded in filaments and that in turn filaments are embedded in walls.  }

\subsection{Cosmic Web dark matter particle and halo population}
\label{subsec:population}
An interesting and important issue for the study of structure formation is our understanding on how far the cosmic structure in the dark matter distribution is reflected in the halo distribution. In \reffig{fig:env_particles} we show the particles in our numerical simulation splitted according to the environment in which they reside. The filament environment correctly traces the largest linear particle concentrations. This can clearly be seen in the prominent filaments. While the thinner structures are less populated with dark matter particles, they are identified as filaments due to their higher local contrast. \MC{\Spider finds that on average filaments have an overdensity $1+\delta$ around 11, as can be seen by comparing the mass and volume filling fractions given in \reftab{tab:filling_fractions}.} On the other hand, walls are much more tenuous structures and this is visible in the particle distribution. When seen edge on, the sheets appear as prominent structures in the particle distribution. But when looked at face on, they are sparsely populated by particles and hence difficult to identify. When comparing the particle distributions in walls and voids there seems to be little difference in the particle densities. This is just a projection effect, with walls having \MC{on average an overdensity around $1.4$ versus an overdensity of $0.2$ for voids (see  \reftab{tab:filling_fractions} for details).}
The success of the structure finding method can be easily assessed by observing that the dark matter particles in void regions do not have any significant structure present in their distribution.

\begin{table}
    \caption{\MC{The mass and volume filling fractions for the environments identified using \Spider. The results presented here are for the larger $200\Mpch$ simulation. The mass and volume filling fractions for the other five methods are given in appendix \ref{sec:optimal_detection_threshold}.}}
    \label{tab:filling_fractions}
    \begin{tabular}{lcc}
        \hline
        Environment & Mass fraction ($\%$) & Volume fraction ($\%$) \\
        \hline
        clusters    & 8.0   & 0.027 \\
        filaments   & 51.3  & 4.35  \\
        walls       & 24.0  & 16.8  \\
        voids       & 16.7  & 78.8  \\
        \hline
    \end{tabular}
\end{table}

\begin{figure}
    \centering
    $\begin{array}{c}
    \includegraphics[width=.65\linewidth,angle=-90.0]{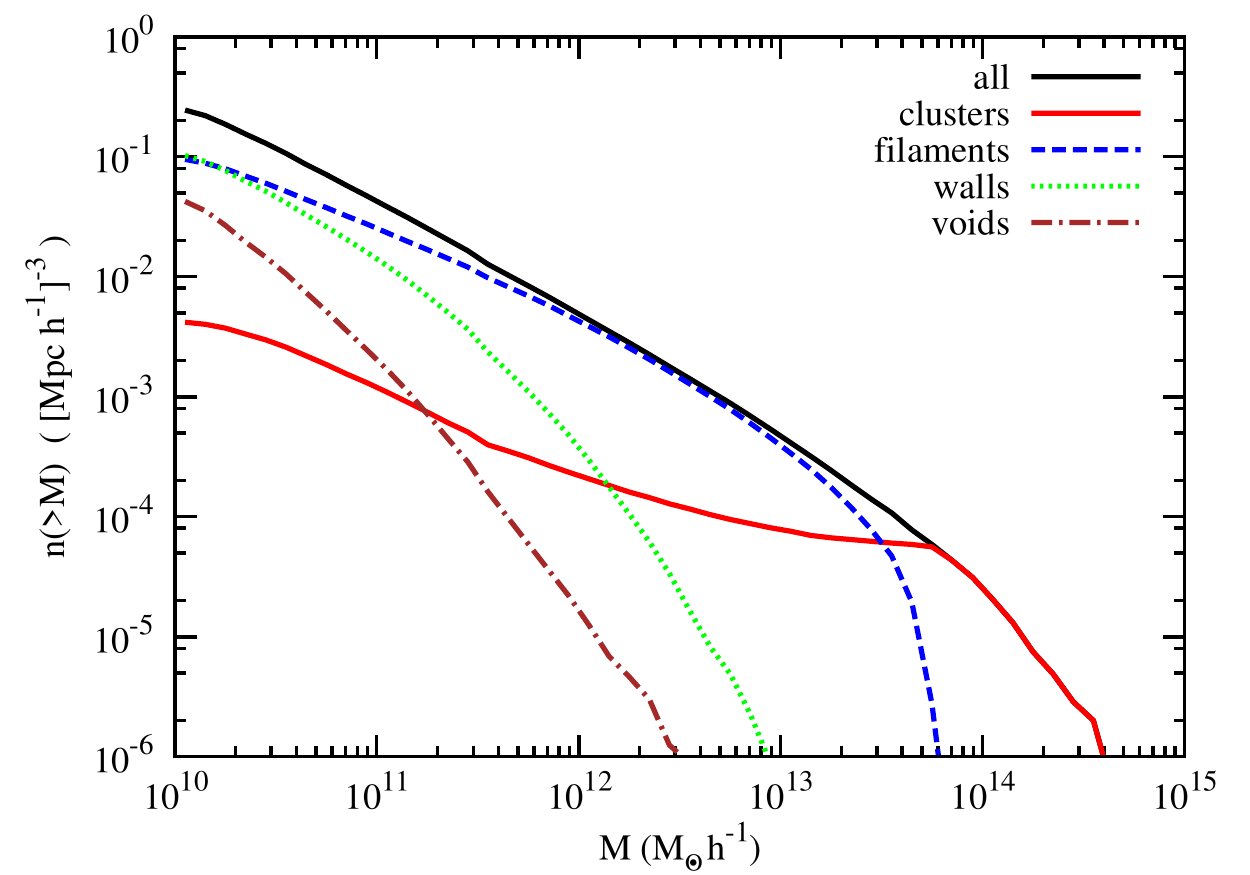} \\
    \includegraphics[width=.65\linewidth,angle=-90.0]{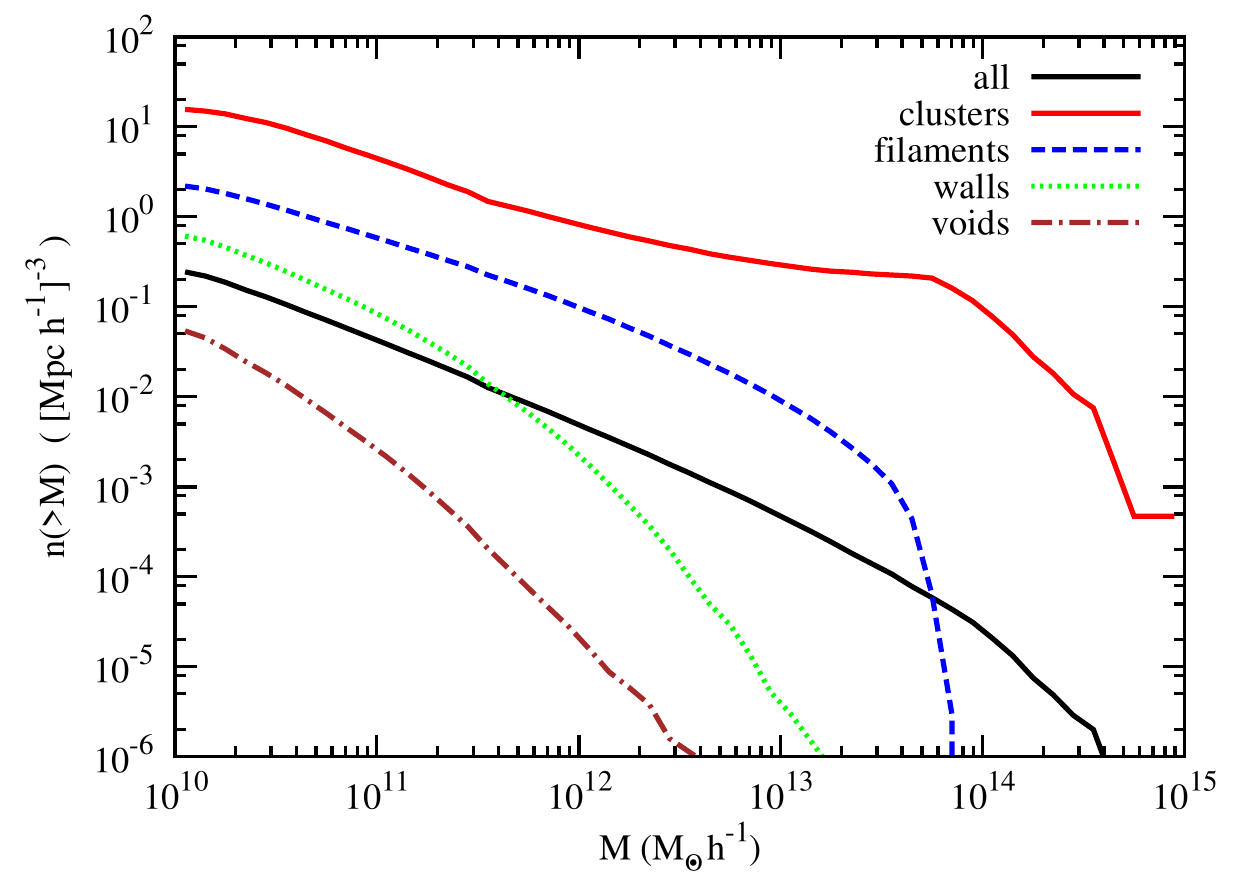}
    \end{array}$
    \caption{\MC{The cumulative halo mass function segmented according to the components of the Cosmic Web as identified by \SpideR. The upper panel gives the cumulative mass function normalized according to the volume of the whole simulation box. The lower panel gives the cumulative mass function normalized according to the volume of each environment (see \reftab{tab:filling_fractions}).}}
    \label{fig:massFunction}
\end{figure}

\MC{A more interesting picture is found when looking at the variation of the dark matter halo populations with the Cosmic Web environment. This is shown in \reffig{fig:env_halos} where the haloes are coloured and scaled according to their mass. \reffig{fig:env_halos}, in conjunction with \reffig{fig:massFunction} which gives the cumulative halo mass function split according to environment, offers a very suggestive picture. We find that all of the massive haloes with $M\ge5\times10^{13}\Msolar$ are located in cluster environments\footnote{Remember that the $5\times10^{13}\Msolar$ mass threshold was introduced as a lower mass cut-off in \refsec{subsec:clusters}. While this lower mass threshold depends on the cluster definition one chooses, the method is very successful in identifying as clusters all the haloes above the cut-off mass.}. Moreover, clusters are the most crowded regions when it comes to haloes of all masses, with halo overdensities about 10 times larger than in filaments and about 100 times larger than in the full simulation box (see lower panel in \reffig{fig:massFunction}).}

\MC{When looking at filaments we see that these environments are also crowded when it comes to haloes. Most of the $10^{12}\Msolar$ and higher mass haloes are located in filaments, while the lower mass objects are a factor of 10 more common in filaments than on average in the universe. From \reffig{fig:env_halos} we see that even the more tenuous filaments have a large number of haloes which shows the power of our method to correctly identify the filamentary environments.}

\MC{The walls are dominated by low mass objects, with the sheets containing only a significant share of the $10^{12}\Msolar$ and lower mass haloes. For a few times $10^{11}\Msolar$ and lower mass haloes, the walls have a similar halo density as the average universe. Compared to voids, the sheets clearly have a much higher halo density and are populated by more massive haloes. In contrast, the voids are very sparsely populated, with extremely few $10^{11}\Msolar$ and higher mass haloes. As in the case of the particle distribution, there does not seem to be any significant structures present in the void halo distribution.}

\subsection{Single scale versus multiscale analysis}
\label{subsec:singleScale}

\begin{figure}
    \centering
    \includegraphics[width=.9\linewidth]{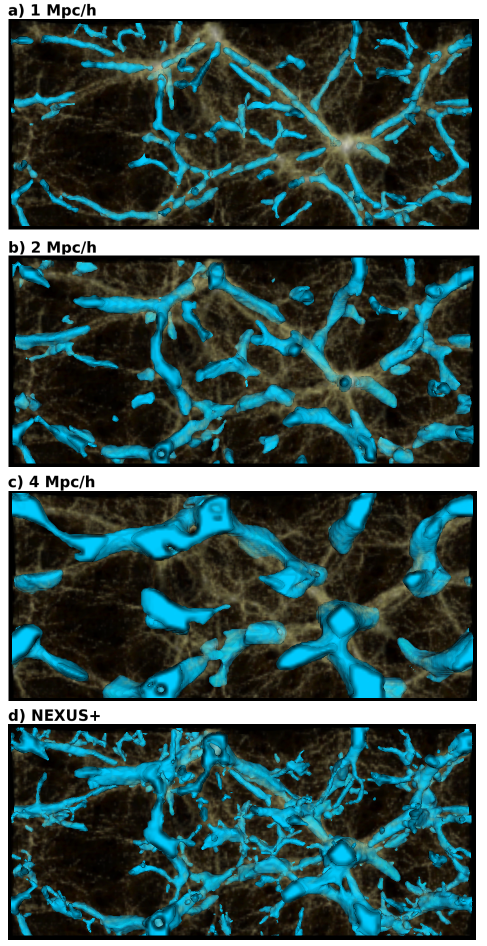}
    \caption{A comparison of the single-scale versus the multiscale approach. Panels a) to c) show the single scale filaments obtained by restricting the \Spider method to the single filter radius of $1$, $2$ and $4\Mpch$ respectively. Panel d) shows the full \Spider results.}
    \label{fig:env_filaments_singleScale}
\end{figure}
\begin{figure}
    \centering
    \includegraphics[width=1.\linewidth]{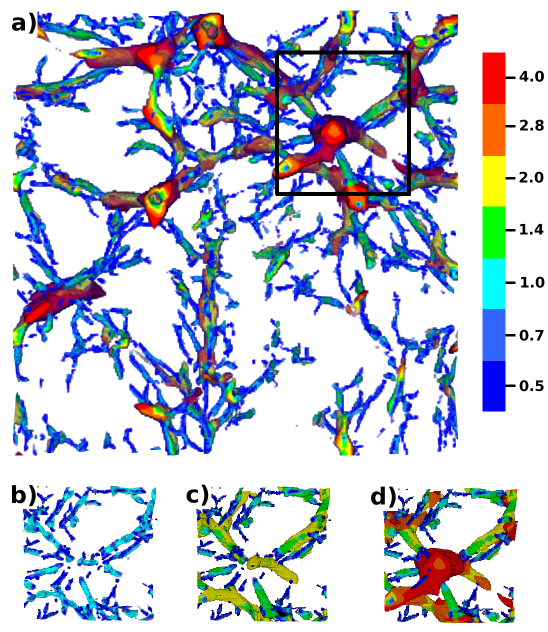}
    \caption{The \Spider filaments coloured according to the smoothing scale that gives the largest filamentary signature for that voxel. The corresponding colour bar indicates the filter scale in units of$\Mpch$. The lower panels b) to d) show a smaller region from the larger volume presented in panel a). These panels show the filaments detected using a maximum of: b) $1\Mpch$, c) $2\Mpch$ and c) $4\Mpch$ smoothing scales.}
    \label{fig:env_scale_space}
\end{figure}

One of the frequent questions that surface when dealing with cosmological structure identification is the optimal value of the smoothing scale \citep[see][]{Hahn2007a,Forero-Romero09}. Often this constitutes a major limitation of single scale approaches. Moreover, until now, the choice of one smoothing length versus another was mostly heuristic. While our multiscale approach does not suffer from these problems, it is certainly a very interesting question to find which scales are the most important and how they compare with results from other works.  

We exemplify the effects of a single scale versus a multiscale approach in \reffig{fig:env_filaments_singleScale}. To obtain the single scale results we restrict the \Spider method to a single filter radius. We then show this result for several values of the filter radius. 
For small smoothing scales, the method detects most of the filaments. While it reproduces correctly the small filaments, it greatly underestimates the thickness of the more prominent objects. As the filter radius is increased, most of the thinner filaments are missed while the larger ones are detected as being thicker and thicker.  
In conclusion, applying a filter of a given radius makes the environment detection method sensitive only to objects similar or larger than that smoothing scale. Moreover, at a given filter radius, all filaments have very similar diameters which are given by the value of the smoothing radius.

By comparing the single-scale results versus the full \Spider results (see \reffig{fig:env_filaments_singleScale} panel d) we conclude that the $2\Mpch$ smoothing scale gives the closest match to the multiscale filaments. However, there are a lot of important differences, with many thin filaments missing in the single scale picture. Another striking difference is the smaller diameters for the larger filaments in the single scale versus the full \Spider results. These results strongly show the need of a multiscale approach to be able to fully trace all the features of the Cosmic Web.

When restricted to a single scale, the \spiderDen and \spiderVeldiv methods show an even stronger difference between results at different filter radii. This trend is so strong that it is very difficult to find a single smoothing scale that matches even remotely the multiscale results. Even the single scale \spiderTidal and \spiderVelshear methods show significant differences between scales, though less prominent than the results in \reffig{fig:env_filaments_singleScale}. The weaker dependence on scale is due to the steep drop at small wavelengths of the gravitational and velocity potentials spectra (see \reffig{fig:input_field_spectra}) which means that large scale modes contribute more than for the other methods. For the single scale \spiderTidal and \spiderVelshear methods we find that $1\Mpch$ smoothing radius results are the closest match to the full multiscale results. This is in contrast with \citet{Hahn2007a}, who argue for a $2.1\Mpch$ filter radius.

Another very interesting question is finding the smoothing scale at which a region has the largest environmental signature. This is illustrated in \reffig{fig:env_scale_space}. We show in different colours the filter radius which gives the strongest filamentary characteristic for a given region. The small filter radii give the strongest signature in the thinner filaments while the larger smoothing scales give a stronger signal for the major filaments. A closer inspection of the prominent filaments shows that their central axes give the strongest filamentary response for small filter radii. Increasing the size of the filter adds larger and larger filamentary regions around this inner central axis. This is visible in panels b) to d) of \reffig{fig:env_scale_space}. This is due to the inner structure of the filaments, with their central axis having a larger density than their periphery.

\section{Comparison to other structure finding algorithms}
\label{sec:comparison}
\begin{figure}
    \centering
    \includegraphics[width=.9\linewidth]{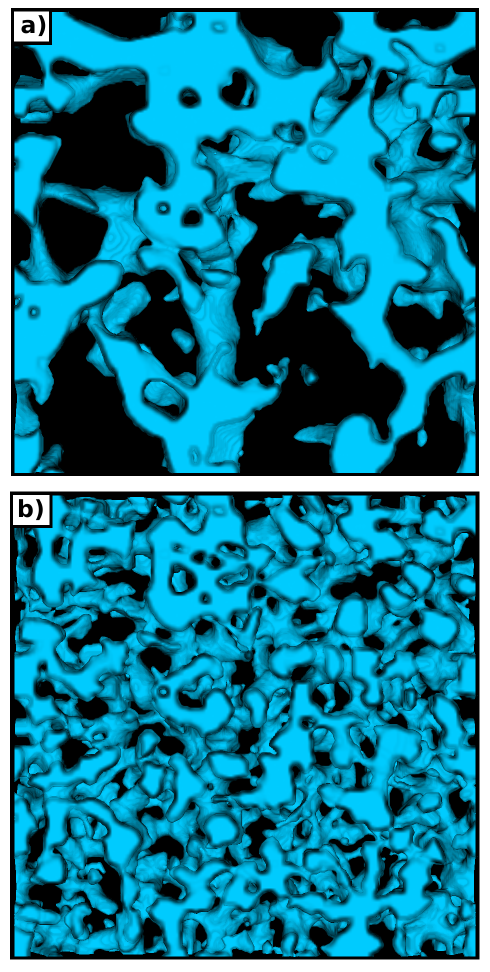}
    \caption{The filamentary network as identified by the algorithms presented in: a) \citet{Hahn2007a} and b) \citet{Zhang09} (see text for details). The results are obtained for a $2\Mpch$ filter width. It shows the same volume of the N-body simulation as in Figures \ref{fig:density_divergence_box} and \ref{fig:env_filaments}.}
    \label{fig:env_noThreshold}
\end{figure}

In this section we discuss the abilities and virtues of the \spider and \Spider algorithms with respect to those of other structure finding algorithms. It will underline the advantages and disadvantages of the instruments described and introduced in this paper. In our discussion we limit ourselves to Hessian and topological based methods. There are a few additional methods that find filaments using directly the particle/galaxy distribution \citep{Stoica05,Stoica07,Stoica10,Chazal09}. While these approaches have the advantage that one does not need to compute the density field, they depend on many free parameters that make their use cumbersome.  

A first group of Cosmic Web identification methods consists of algorithms that use the Hessian of the density or gravitational potential. Both the \spider and \Spider algorithms are part of this class. 

\citet{Hahn2007a} (referred to as HPCD) proposed the use of the tidal field eigenvalues for environment classification\footnote{This method can be easily implemented as a special case of our \spiderTidal algorithm.}. The method uses the criterion that: all positive eigenvalues identify clusters, one negative eigenvalue corresponds to filaments while two negative eigenvalues trace sheets. The result of this method applied to our simulation is shown in \reffig{fig:env_noThreshold}. It is immediately clear that while the inner region of the \hahn results correspond to \spiderTidal (compare to panel b in \reffig{fig:env_filaments}), the \hahn filaments extend to much larger diameters, encompassing substantial parts of wall and void regions. This leads to a large cross-contamination of the Cosmic Web components. A similar method to \hahn, but using the density Hessian instead of the tidal field, was used by \citet{Zhang09}. From \reffig{fig:env_noThreshold}, panel b, is obvious that this last method is not very successful. It leads to even more misclassified regions than the \hahn method. There are two main shortcomings of the two methods: use of a single scale approach and the absence of a threshold to distinguish between significant and spurious detections.

The absence of a detection threshold in the \hahn method was pointed out in \cite{Forero-Romero09}. They suggested that an eigenvalue threshold in the range $0.2-0.4$ gives results in agreement with the visual impression of the Cosmic Web. Using \spiderTidal restricted to the $2\Mpch$ filter radius we do obtain that the filament and wall threshold is 0.42 and 0.2 respectively. While we do confirm the results of \cite{Forero-Romero09}, we stress that \spiderTidal has the advantage of a multiscale approach and does not employ user dependent arguments for specifying the detection threshold.

It is the multiscale character of \spider and \Spider which are crucial for their successful analysis of emerging structures and patterns in the hierarchically evolving cosmic mass distribution. The first version of the multiscale formalism that we have developed into \spider and \Spider is the Multiscale Morphology Filter (MMF), described and introduced in \cite{Aragon07b}. \spider is an extension of MMF to a more physical and versatile algorithm. While MMF is restricted to the density field, the \spider method incorporates, among others, the tidal and velocity fields. Both MMF and \spiderDen find the same filamentary structures, while the second method gives a much better identification of walls (for details see Appendix \ref{sec:optimal_detection_threshold}). More importantly, \Spider seems optimal at capturing the structural intricacies of the Cosmic Web, and thus represents an important advancement within the context of the multiscale scale-space formalism that we have been developing. 

An additional important class of structure identification procedures are based on topological considerations, in principal following an analysis of the Morse-Smale complex of the density field \citep{Sousbie08a,Sousbie2011a,Aragon10}. Given the fact that in general, cosmic density fields behave like a proper Morse function, the assumption is that filaments and walls in the mass distribution can be identified with the manifolds in the density field connecting maxima via saddle points of the field \citet{Sousbie2011a}. Filaments are identified with the line connecting two maxima via a saddle point, and walls with the sheet separating the regions around two minima and centered around a saddle point. \citet{Aragon10} developed a similar strategy by first delineating the watershed basins around the mimima in the field, following the WVF procedure introduced by \cite{Platen07} in which the watershed basins are identified with the voids in the cosmic mass distribution. 
The topological character of the boundary is determined locally via the number of touching watershed basins. Walls are the 2-d manifolds separating two watershed basins, the filaments are their boundaries and identified as the locations where three basins touch each other. 

\MC{\spider and \Spider do find that the largest filaments are between massive density peaks, which is in agreement with the implicit assumptions of the topological methods. But on top of the prominent structures, we also find, especially in the \Spider results, an important network of thinner objects which branches into voids.} These tenuous objects contradict the hypothesis that filaments are always located between density maxima. We also find that walls are not fully continuous sheets and that they sometimes stop as they branch into lower density regions. Thus, the voids fully percolate, with large regions without a clear boundary between adjacent voids. Also, we should note that while the topological based methods can detect the central axis of filaments and inner plane of walls, they cannot assign a natural thickness to the structures. As our results show, both the filament and wall environments have a wide range of sizes. This makes it rather challenging for the topological methods to detect and outline in a natural fashion the regions belonging to a filament around the identified central filament axis.

\section{Conclusions and Prospects}
\label{sec:conclusions}
This work presents the \spider and \Spider methods, which are multiscale and automatic algorithms used for the segmentation of the Cosmic Web into its distinct components: clusters, filaments, walls and voids. We have shown that the environments identified with the two methods correspond very well to the structures visible in the density and velocity divergence fields as well as in the dark matter particle and halo distributions. The success of the method lies in two important ingredients: the use of a scale free approach that makes sure that the algorithm detects structures of all sizes, and the use of a physically motivated threshold to distinguish valid environments from spurious detections. Another strength of the two algorithms is that they do not depend on user set parameters and therefore can be easily applied in a consistent way to multiple data sets.

We have extended the \spider method to detect the Cosmic Web as traced by the density (\spiderDeN), tidal field (\spiderTidaL), velocity divergence (\spiderVeldiV) and velocity shear fields (\spiderVelsheaR). We find that \spiderDen and \spiderTidal are very efficient in identifying the cluster regions when compared with the most massive haloes present in the simulation. The methods perform similarly in the detection of filament and wall regions, with all the prominent structures detected by all methods. The only differences arise in the identification of the more tenuous structures. We find that \Spider performs better in tracing the weaker filaments and walls.

From all the methods presented in this paper, we find that \Spider is the most successful one in tracing the Cosmic Web. Its main advantage comes from the use of the \logFilter filter which is designed to better deal with the orders of magnitude difference in the density field between low and high density regions. The filamentary and wall like environments detected with \Spider contain complex networks of prominent structures that branch out into more tenuous ones until they finally disappear out into underdense regions.

We showed in \refsec{sec:comparison} that \spider and \Spider have several advantages compared to other Hessian based methods: the use of a multiscale approach and a physically motivated threshold for identifying the significant environments. Compared to topological methods, our tools are able to detect the filamentary/wall regions and not only their central axis/plane. Moreover, we do not make the assumptions that filaments extend between density maxima and walls separate density minima basins. Our results show that these assumptions do not always hold.

Equipped with the \spider and \Spider methods, we plan to address a range of cosmological issues. The ability to identify filaments and walls over a range of scales, in both numerical and observational datasets, allows us to study not only environmental factors affecting the formation and evolution of dark matter haloes and galaxies, but also the hierarchical buildup of the Cosmic Web itself. Our first priority lies with a systematic study of environmental factors affecting the evolution of galaxies. The fact that we can identify galaxies and haloes within finely outlined filaments and sheets will allow us to determine which physical characteristics are most sensitive to the environment, which galaxies and haloes are most sensitive to large scale influences, and to study the causes that give rise to this dependency. 

A major point of our interest is performing a systematic comparison between the structures traced by the density, velocity and tidal fields in order to understand which of these physical influences are most decisive in determining the global outline of the Cosmic Web. In addition to our current focus on the large scale structure of the dark matter distribution, we will also direct our study to the structure of the gaseous Cosmic Web. Comparison of the IGM with the dark matter structures in numerical simulations will be instrumental in understanding how the Cosmic Web can be traced both in the galaxy and cosmic gas distribution. This will be essential for relating the distribution of HI in the local Universe to the overall large scale structure found in the galaxy distribution \citep{Popping11}, and will help understand recent findings such as a small-scale HI filament in a void \MC{(Beygu et al., in prep.)}.

Finally, the \spider and \Spider procedures are perfectly suited for a systematic appraisal of the structures found 
in maps produced by galaxy redshift surveys such as SDSS and 2MRS.

\section*{Acknowledgement}
We are grateful to Miguel Aragon-Calvo for useful discussions about the MMF method and for his early IDL implementation. We also thank Wojciech Hellwing and Carlos Frenk for inspiring discussions and suggestions. \MC{We are grateful to the anonymous referee whose comments and questions improved the presentation of this paper.} All simulations and analysis were performed on the Gemini machines at the Kapteyn Astronomical Institute, Netherlands.

\newcommand{\jcap}{Journal of Cosmology and Astroparticle Physics}
\bibliographystyle{mn2e}
\bibliography{NEXUS_bib}

\appendix
\section{Optimal filament and wall detection}
\label{sec:optimal_detection_threshold}
This section deals with identifying the signature threshold used for the detection of the Cosmic Web filaments and walls. When applying the environment detection algorithm every region of space is assigned an environment signature $\mathcal{S}$ (for definition see \eq{eq:response}). A large signature corresponds to very prominent structures while a zero or small one corresponds to null detections. Since many regions of space will have a signature value between the two extremes, we need to identify a signature threshold that differentiates between valid structures and spurious detections. All regions with signatures larger than the threshold will correspond to valid environments.

The simplest way to define the Cosmic Web is using the tidal field, since this is what drives the anisotropic collapse. Using the eigenvalues of the tidal tensor one defines filaments and walls as regions with one and two negative eigenvalues \citep[see][]{Hahn2007a}. The major problem with this approach is that it gives unrealistic looking environments, with only a small fraction of the volume ($\sim10\%$) occupied by voids -- this is in stark contrast with the observational data.
\begin{figure}
    \centering
    \includegraphics[width=0.85\linewidth]{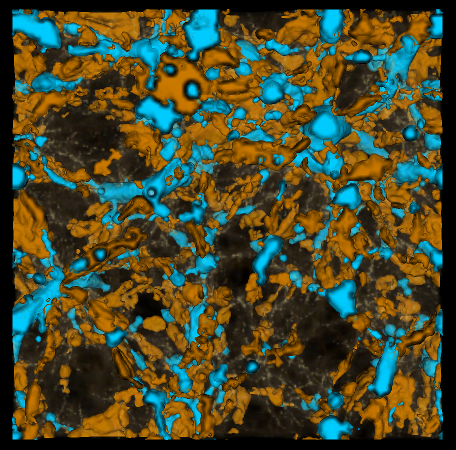}
    \caption{A 3D volume rendering of the \spiderDen filaments (blue) and walls (orange) when the detection threshold is taken as the percolation point.}
    \label{fig:env_percolation_threshold}
\end{figure}
\begin{figure}
    \centering
    \includegraphics[width=0.65\linewidth,angle=-90.0]{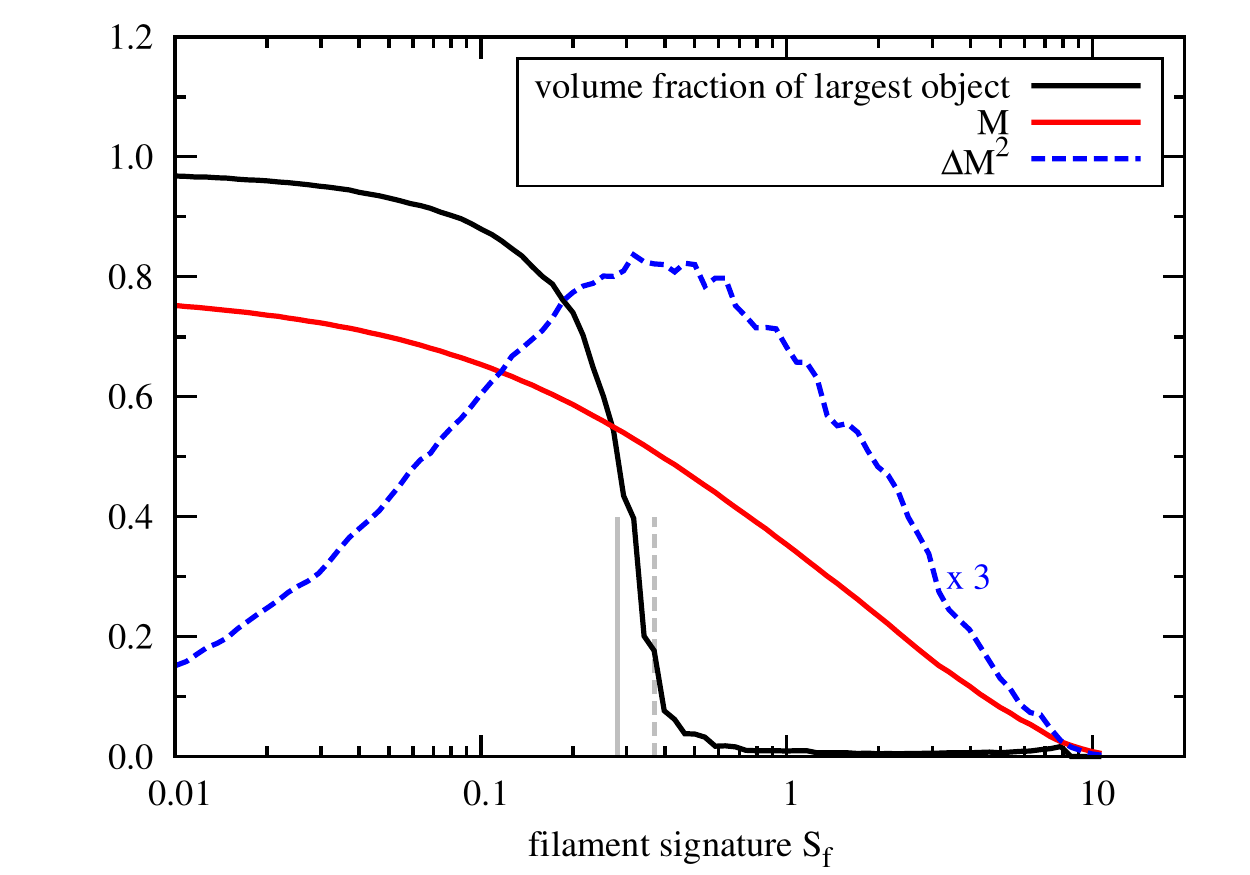} \\
    \includegraphics[width=0.65\linewidth,angle=-90.0]{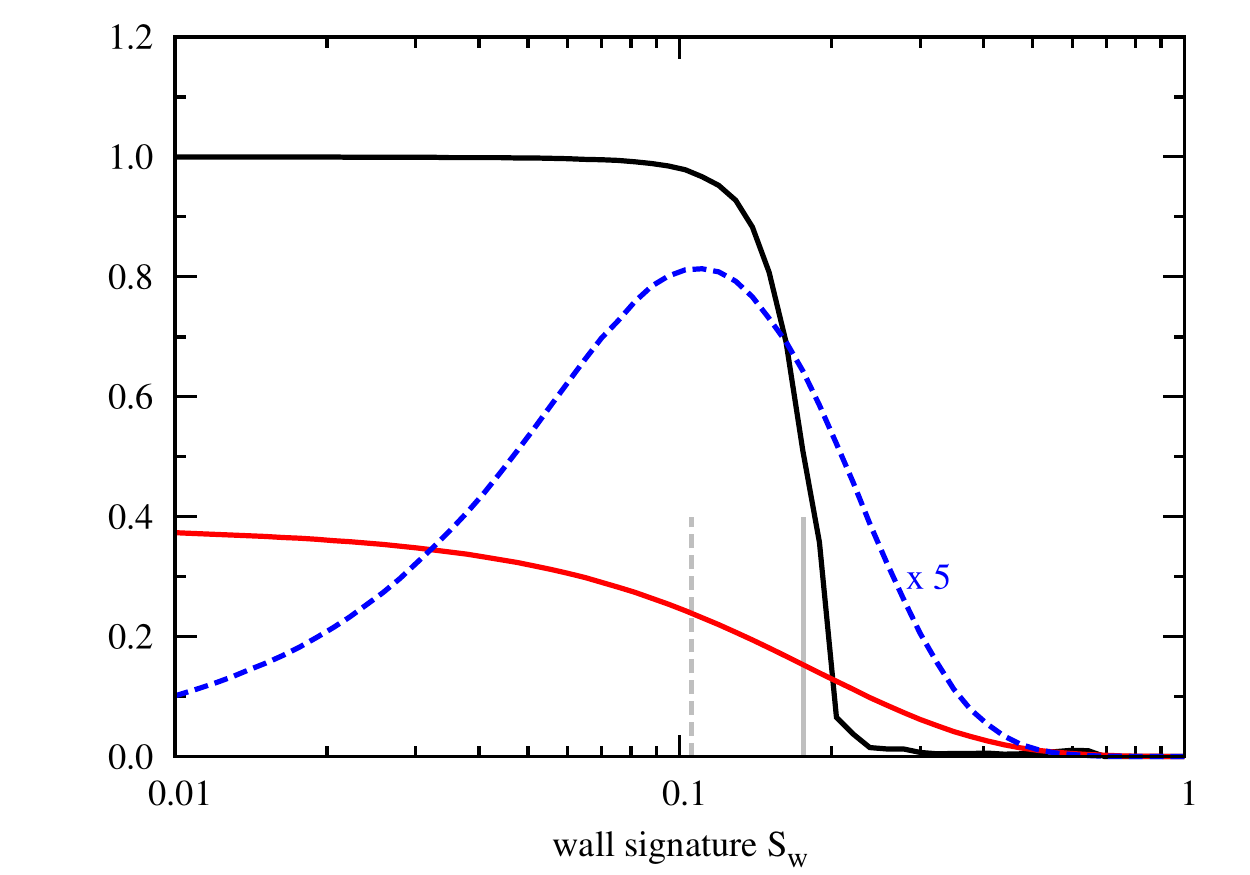}
    \caption{The dependence of the mass fraction in the components of the Cosmic Web as a function of environment signature. The two panels shows the dependence for filaments and respectively walls obtained using the \spiderDen method. The continuous red curve gives the mass fraction $M$ in filaments/walls while the dashed blue curve gives $\Delta M^2$ (see text for definition). The black curve gives the volume fraction of the largest filament/wall. The sharp transition of this curve from 0 to 1 marks the percolation threshold. The vertical gray lines mark the percolation point (filled line) and the peak of $\Delta M^2$ (dashed line).}
    \label{fig:appendix_optimalThreshold}
\end{figure}

A second approach is to detect only the most significant filaments and walls. This is the procedure we follow in this paper. \cite{Aragon07b} have argued that the percolation threshold of filaments/walls offers a natural way of identifying the prominent structures. We find that indeed the percolation threshold gives a good identification of the filamentary network, but fails in detecting the sheets. The walls obtained via this method are made of many small patches that do not show the large scale cohesion expected for void boundaries (see \reffig{fig:env_percolation_threshold}). Another downside is the dependence of the percolation threshold on the grid resolution used to analyse the data. But more importantly for this work, many large differences between the results of the 6 methods described here can be attributed to different percolation properties of the filamentary/wall networks in each method. We analyse this in more details later on.
\begin{figure}
    \centering
    \includegraphics[width=0.65\linewidth,angle=-90.0]{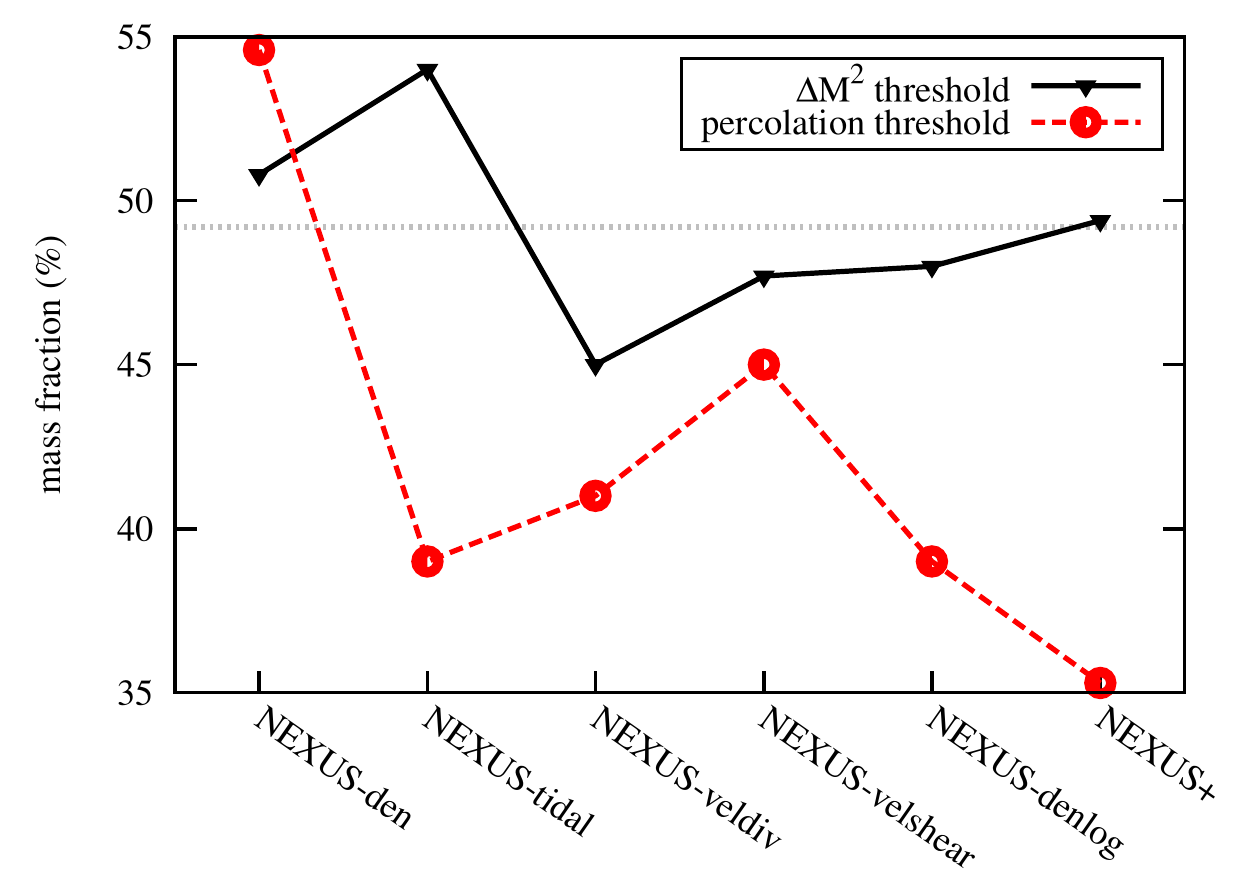} \\
    \includegraphics[width=0.65\linewidth,angle=-90.0]{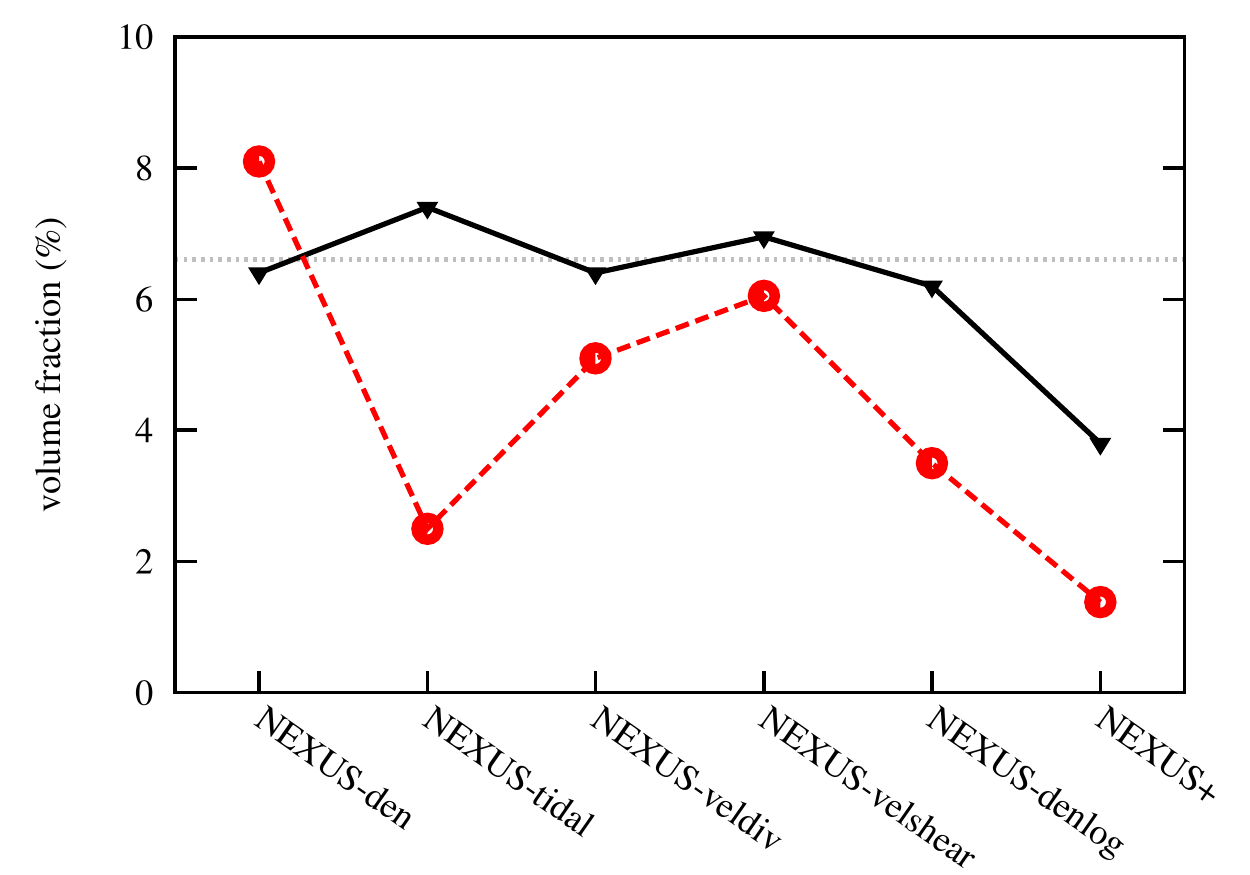}
    \caption{Comparison of the mass fraction (upper panel) and volume fraction (lower panel) in filaments detected using the 6 methods described in this paper. The continuous black line describes the filaments detected using the $\Delta M^2$ threshold method (described in this section) while the dashed red line shows the filaments detected using the percolation threshold \citep{Aragon07b}. The horizontal dotted line give the average mass and volume fraction for the $\Delta M^2$ results.}
    \label{fig:appendix_percolationComparison}
\end{figure}

Most of the mass in filaments and walls is given by regions with a narrow range in environment signatures. This is illustrated by the red curve of \reffig{fig:appendix_optimalThreshold} which gives the mass fraction $M$ in filament/wall regions as a function of the signature threshold $S$. The rapid increase in the filament/wall mass can be appreciated when computing the mass change:
\begin{equation}
    \Delta M^2 = \left| \frac{dM^2}{d\log\mathcal{S}} \right|.
\end{equation}
This is represented by the dashed blue curve in \reffig{fig:appendix_optimalThreshold}. We found that the peak position in $\Delta M^2$ gives a robust and natural way of identifying the most significant filaments and walls. The environments detected using the $\Delta M^2$ peak threshold are shown in Figures \ref{fig:env_filaments} and \ref{fig:env_walls}. This new threshold captures very well the filamentary and wall features seen in the density and velocity divergence fields from \reffig{fig:density_divergence_box}. Moreover the peak of $\Delta M^2$ for the \spiderDen filaments is very close to the percolation threshold and hence reproduces the success of using percolation as a good tracer of the filamentary network\footnote{The percolation threshold and the $\Delta M^2$ peak are similar only for the \spiderDen filaments, in general there are large offsets between the two values. This was expected since \cite{Aragon07b} have shown the success of the percolation threshold for filament detection only when using the density field as tracer of Cosmic Web environments.}.

There are two major improvements when using the $\Delta M^2$ peak versus the percolation point as the detection threshold for the most significant environments. The walls detected via the $\Delta M^2$ method have a bigger large scale cohesion than the results of the percolation method. This can clearly be seen by comparing the \spiderDen walls from \reffig{fig:env_walls} and the percolation walls in \reffig{fig:env_percolation_threshold}. The second enhancement comes from a more robust detection threshold. This can be seen in \reffig{fig:appendix_percolationComparison} where we compare the mass and volume fraction in filaments detected using the 6 methods introduced in this paper. Notice the large variation in mass fraction between the \spiderDen and \Spider results for the percolation method -- with the former having almost twice as much mass than the latter one. In the case of the $\Delta M^2$ threshold all methods give a similar mass fraction, with a much smaller scatter around the mean. The same behaviour can be seen in the volume fraction plot, where the $\Delta M^2$ threshold gives more consistent values. The only exception is the \Spider volume fraction whose lower value is due to the method itself and not the detection threshold used (see \refsec{subsec:filaments} for more details).

\end{document}